\documentclass[12pt]{article}

\usepackage{fancyhdr}
\usepackage{fullpage}
\usepackage{amsmath}
\usepackage{amsbsy}
\usepackage{amssymb}
\usepackage{amscd}
\usepackage{amsfonts}
\usepackage{graphics}
\usepackage{verbatim}
\usepackage{epsfig}
\usepackage{xspace}
\usepackage{euscript}
\usepackage{alltt}
\usepackage{multirow}
\usepackage{float}
\usepackage[colorlinks]{hyperref}
\usepackage{color}
\usepackage{t1enc}
\usepackage{times,exscale}
\usepackage{graphicx,calc}
\usepackage{subfig}
\usepackage{epstopdf}
\usepackage{verbatim}
\usepackage{geometry}
\usepackage{bm}
\usepackage{caption}

\title{Strain engineering of doped hydrogen passivated silicon quantum dots}
\begin{document}
\maketitle
\begin{center}

{{Swarnava Ghosh$^{a*}$ and Markus Eisenbach$^a$}} \footnote{
Notice: This manuscript has been coauthored by UT-Battelle, LLC, under Contract No. DE-AC0500OR22725 with
the U.S. Department of Energy. The United States Government retains and the publisher, by accepting the article for
publication, acknowledges that the United States Government retains a non-exclusive, paid-up, irrevocable, world-wide
license to publish or reproduce the published form of this manuscript, or allow others to do so, for the United States
Government purposes. The Department of Energy will provide public access to these results of federally sponsored
research in accordance with the DOE Public Access Plan (\href{http://energy.gov/downloads/doe-public-access-plan}{http://energy.gov/downloads/doe-public-access-plan}).}

{$^a$ National Center for Computational Sciences, Oak Ridge National Laboratory, Oak Ridge, TN, 37830}\\
{$^{*}$ Email: ghoshs@ornl.gov}

\end{center}
\begin{abstract}
Silicon quantum dots are nanomaterials that are attractive candidates for photovoltaic applications. Doping of these materials creates p-n junctions and is important for solar cells. In this work, we present a first-principles study of the coupled influence of doping and strain on the stability, energy gap, Fermi level, electronic density, and density of states of hydrogen-passivated silicon quantum dots. We find that the cohesive energy and the energy gap decrease with increasing quantum dot size and are strongly influenced by strain. Furthermore, the response to strain also depends on the size of the quantum dot and dopant type. We present expressions of cohesive energy and energy gap as power-law of size and polynomial dependence on strain. We also show that the Fermi energy increases with size for pristine and p-type doping but decreases with size for n-type doping. We also discuss the influence of strain and dopant type on the density of states and electron density of the quantum dots.
\end{abstract}
\section{Introduction}
Nanoclusters are single or polycrystalline materials with dimensions less than 100 nanometers. An interesting phenomenon in nanoclusters is the spatial confinement of electron-hole pairs in one or more dimensions, also known as quantum confinement. This phenomenon is typically observed when the dimension of the nanocluster is less than twice the Bohr radius of the average elemental composition of the nanocluster \cite{soga:2006,wheeler:2013}, but this effect is weak in nanoclusters with diameters twice the average Bohr radius. These small nanoclusters{,} which exhibit quantum confinement{,} are also called quantum dots. For silicon, the Bohr radius is 5 nanometers, and hence silicon nanoclusters of diameter less than 10 nanometers display quantum confinement \cite{wheeler:2013,warner:2005,cho:2007}. Semiconducting quantum dots show fascinating properties that are not observed in the bulk. A well-known example of this is the photoluminescence of silicon quantum dots below a critical diameter due to strong quantum confinement\cite{peng:2006,canham:1990,delley:1993,wang:1994,Proot:1992,delerue:2000}. In contrast, bulk silicon does not emit visible light. Surface passivation of silicon quantum dots by hydrogen increases {their} stability \cite{peng:2006,ougut:1997}.

Nanoclusters are attractive candidates for solar cell applications because they have tunable bandgaps. These can be classified into compound semiconductor quantum dots, such as PbS, CdSe \cite{duan:2015}, and single-element quantum dots such as germanium and silicon. Silicon quantum dots have the potential to be integrated into existing silicon solar cells to increase their efficiency \cite{oliva:2016}, and is thus of technological relevance \cite{oliva:2016,pi:2011,pi:2012,huang:2012,hao:2009,hao:2009b,hao:2008,perez:2009,pagliaro:2008,sze:2008}. Doping silicon quantum dots by p-type, such as elements from Group 13 and n-type, such as elements from group 15 creates p-n junctions and is important for solar cell applications. Typical p-type dopants are boron{,} and n-type dopants include phosphorous.

Doping alters the properties of quantum dots \cite{ossicini:2005,pi:2012,oliva:2016}. Typically P atoms are used as n-type dopants and B atoms as p-type dopants for silicon. The location of dopants in a silicon nanocluster can affect the electrical activity and optical properties of the nanocluster \cite{pi:2012}. Thermodynamically, surface doping is more stable than interior doping \cite{pi:2012}. Simultaneously doping silicon quantum dots with P (n-type) and B (p-type) results in the carriers being perfectly compensated, leading to smaller formation energies than single-doped cases \cite{ossicini:2005}. Structural distortion around impurities is also observed in these co-doped silicon quantum dots, and the energy gap of the co-doped quantum dot is smaller than the energy gap of the pure quantum dot \cite{ossicini:2005}. Doped silicon quantum dots can also be used as a plasmonic material with infrared optical resonances{,} and the plasmon resonance can be tuned by varying the dopant density \cite{zhang:2017}. These results show that dopant engineering can carefully tune the {photoluminescent} properties of these quantum dots.

The nature of doping can be categorized into three types \cite{oliva:2016}. In the first type, the dopant atom is introduced within the crystal lattice of the nanocluster and is located in the interior. For the second type, the dopant atom is at the surface of the nanocluster. For the third type, the dopant atom is at the interface between the nanoparticle and the surrounding matrix. The reader is referred to a comprehensive review \cite{oliva:2016} of the various synthesis methods for doping silicon quantum dots. 

Density Functional Theory studies show that energy gaps of hydrogen-passivated silicon clusters, wires, and slabs depend on size \cite{delley:1995} and that strain influences the energy gap of silicon quantum dots \cite{peng:2006}. Furthermore, these quantum dots can be categorized into three regimes according to their energy gap variation with strain. For small quantum dots of sizes less than 1 nanometer, the energy gap decreases linearly with strain, however, the reverse trend is observed for quantum dots of sizes greater than 2 nanometers. Interestingly, strain shows negligible influence on the energy gap in quantum dots with diameters ranging from 1-2 nanometers. Calculations of quasiparticle gaps, self-energy corrections, exciton Coulomb energies{,} and optical gaps in hydrogen passivated silicon quantum dots show that self-energy correction is size dependent and decreases with an increase in quantum dot size \cite{ougut:1997}. For silicon wires of diameters up to 1.5 nanometers, calculations also show that quantum confinement and surface effects are responsible for luminescence in porous silicon \cite{buda:1992}. Interestingly, a study of germanium quantum dots shows that the absorption spectra show bulk-like character for cluster sizes of 250 atoms or more \cite{melnikov:2003}. 

As discussed above, research has focused on separately studying the role of strain and doping quantum dots{;} a study to account for size, strain{,} and simultaneously doping of silicon quantum dots from first principles is lacking. In this work, we report a first-principles study of the simultaneous influence of doping, size, and strain on the stability, energy gap, Fermi level,  density of states, and electronic density of hydrogen-passivated silicon quantum dots. 

\section{{Methods}}

\begin{figure}[H]\centering
\includegraphics[keepaspectratio=true,width=0.8\textwidth]{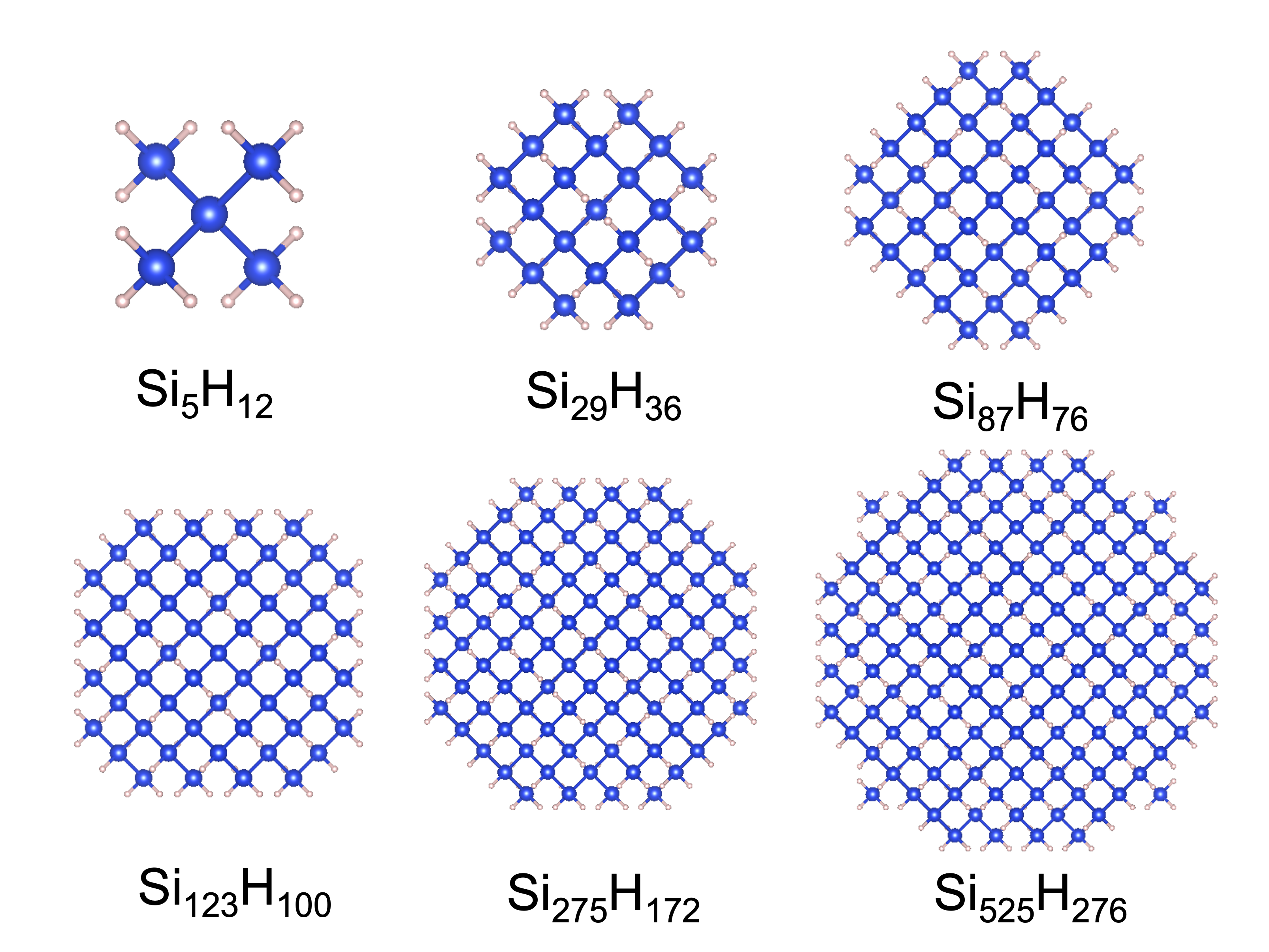}
\caption{Atomic structures of silicon quantum dots passivated with hydrogen considered in our study. {The atoms shown are projected from a three-dimensional space to a two-dimensional space}} \label{Fig:Sinano}
\end{figure}

We discuss the variation of cohesive energy, energy gap, and Fermi energy, with quantum dot size for pristine and doped hydrogen passivated silicon quantum dots. The three-dimensional quantum dots are created by spherically terminating a diamond cubic bulk silicon structure and then attaching H atoms at the dangling bonds on the surface. The quantum dots are doped by substituting a silicon atom at the center or the edge {with} a dopant. In the first case, the dopant is at the center, and the dopant atom is bonded to four silicon atoms, and in the second case, the dopant atom is placed at the edge and is bonded to two silicon atoms and two hydrogen atoms. In our work, we consider quantum dots of size between 0.63 nanometers with chemical formula Si$_5$H$_{12}$, to 2.97 nanometers with chemical formula Si$_{525}$H$_{276}$. The structures of these quantum dots are depicted in Figure \ref{Fig:Sinano}. We consider p-type doping with B and Al atoms and n-type doping with P and As atoms. We apply strain by isotropically displacing the atom positions of the geometry-optimized unstrained quantum dot. The strain considered in this work is between -8\% (compression) to 8\% (tension).

We performed Kohn-Sham Density Functional Theory \cite{hohenberg:1964,kohn:1965} simulations using the Simulation Package for Ab-initio Real Space Calculations (SPARC) framework \cite{ghosh:2017a,ghosh:2017b}, and employed the isolated cluster boundary conditions with {a} minimum distance between an atom and the cell boundary as 12 Bohr. We use twelfth-order accurate finite difference approximation {,} and the real-space mesh spacing was chosen to be 0.5 Bohr. The smearing is used as $1 \times 10^{-3}$ Ha. We use norm-conserving Troullier Martins pseudopotentials \cite{troullier:1991} and the Perdew–Wang parametrization \cite{perdew:1992} of the Local Density Approximation (LDA) correlation energy calculated by Ceperley–Alder \cite{ceperley:1980}. The calculated equilibrium lattice constant for silicon is 10.157 Bohr. To calculate the equilibrium geometry, we relax the cell and the atom positions such that the atomic forces are below 0.01 eV/Angstrom. 

\section{Effect of size, doping, and strain on properties}

\subsection{Cohesive energy}\label{Sec:CE}
The cohesive energy of a quantum dot with the chemical formula Si$_{n-1}$H$_{m}$X, where X=P, B, As, and Al is
\begin{equation}
E_{coh} = E_{Si_{n-1}H_{m}X} - (n-1)E_{Si} - mE_H - E_X \,\,,
\end{equation}
where $E_{Si_{n-1}H_{m}X}$ is the energy of the quantum dot, $E_{Si}$, $E_H$ and $E_{X}$ are the energies of isolated silicon, hydrogen and a dopant atom (P, B, As, and Al), respectively. Cohesive energy is used to understand the stability of quantum dots. A negative value of cohesive energy implies the quantum dot is energetically preferred over its individual isolated atoms, and hence stable. 

\begin{figure}[H]\centering
\subfloat[center]{\includegraphics[keepaspectratio=true,width=0.5\textwidth]{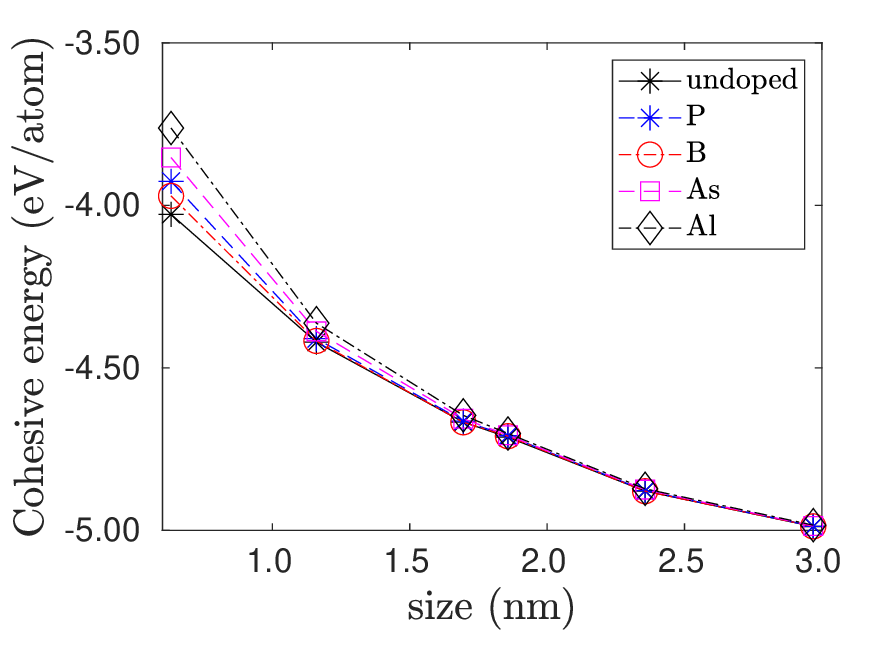}\label{Fig:R:CE}}
\subfloat[edge]{\includegraphics[keepaspectratio=true,width=0.5\textwidth]{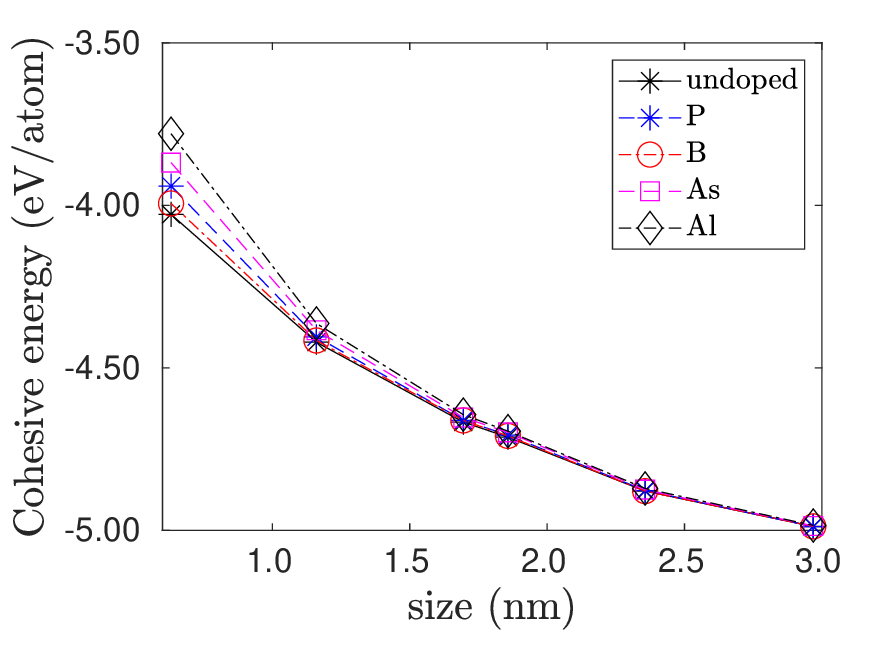}\label{Fig:R:CE:Edge}}
{\caption{Varation of cohesive energy of doped silicon quantum dots with quantum dot size. No strain is applied {to} the quantum dots. In (a) the dopant atom is placed at the center of the quantum dot{,} and in (b) the dopant atom is placed at the edge of the quantum dot.}\label{Fig:CE}}
\end{figure}

\begin{figure}[H]\centering
\subfloat[center]{\includegraphics[keepaspectratio=true,width=0.5\textwidth]{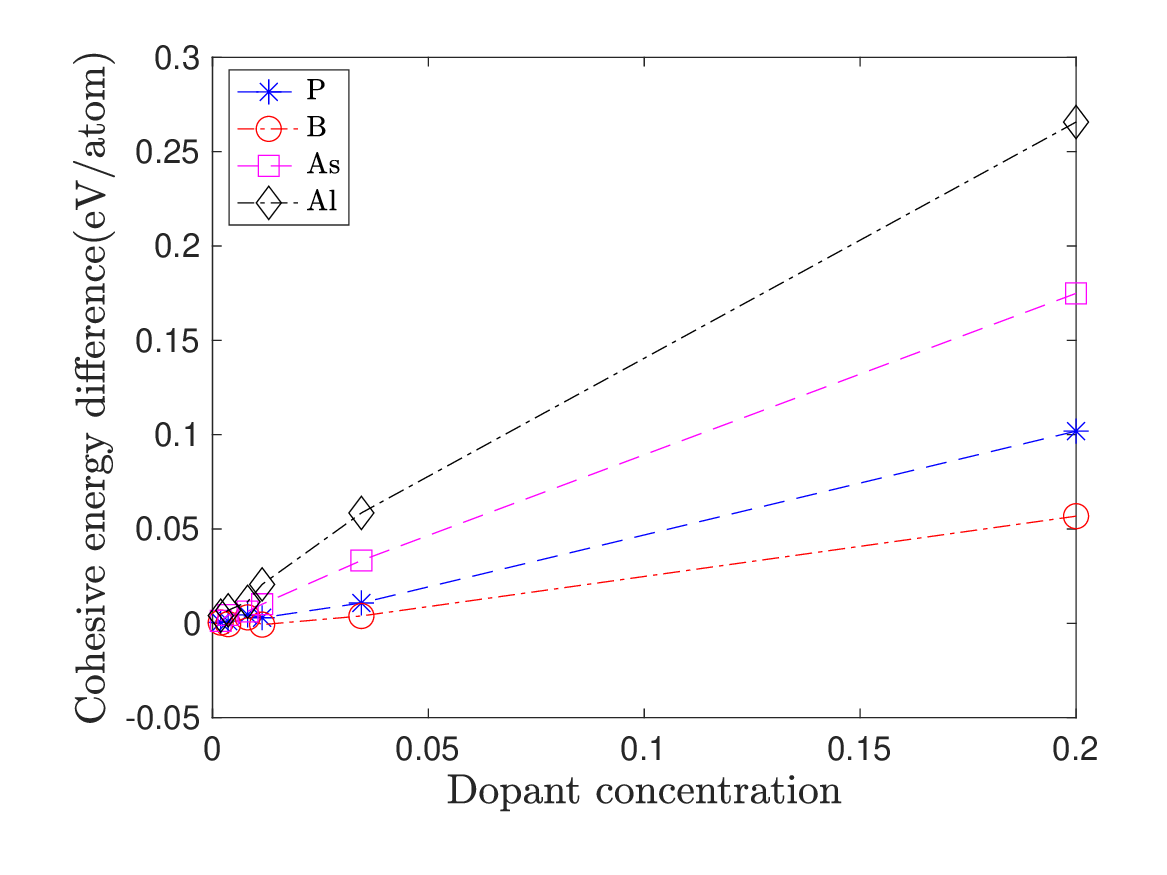}\label{Fig:C:CE}}
\subfloat[edge]{\includegraphics[keepaspectratio=true,width=0.5\textwidth]{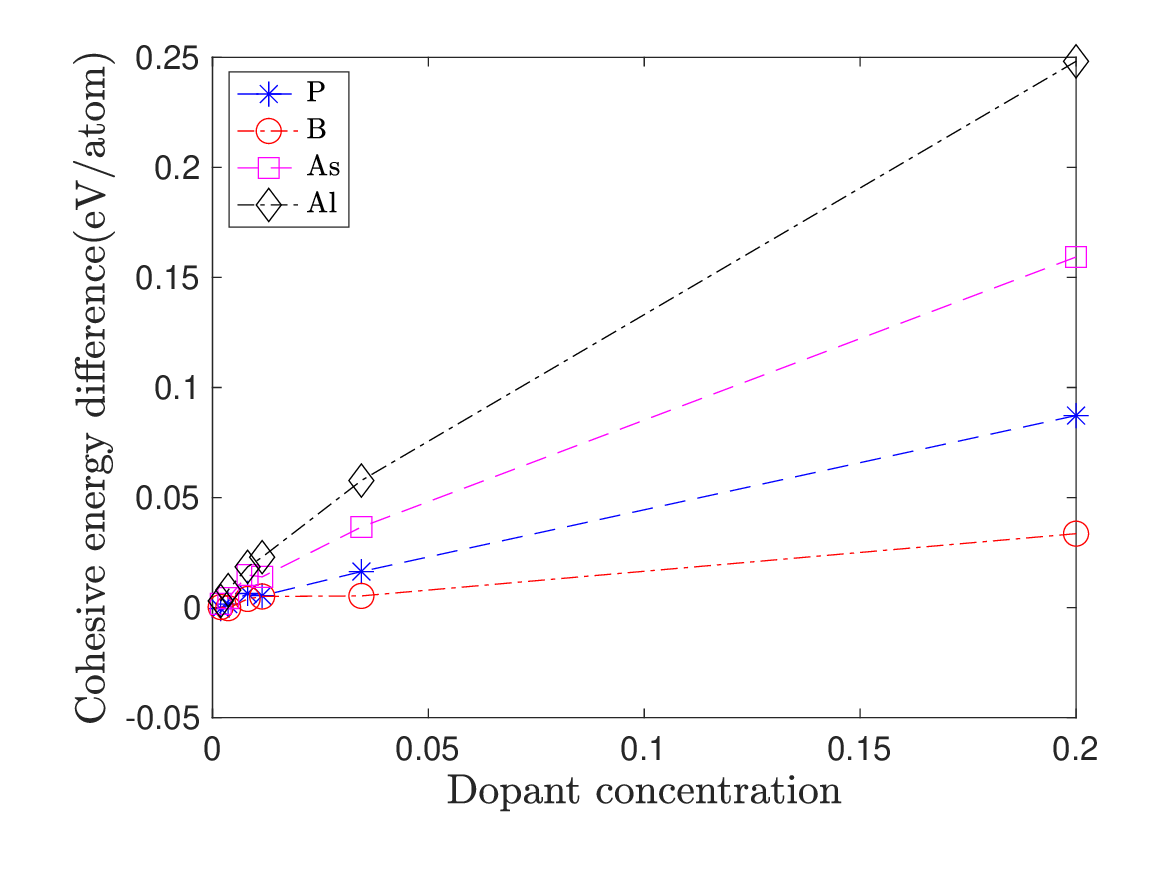}\label{Fig:C:CE:Edge}}
{\caption{Variation in the difference in cohesive energy of doped silicon quantum dots with dopant concentration. The difference is calculated with respect to the undoped case. No strain is applied {to} the quantum dots. In (a) the dopant atom is placed at the center of the quantum dot and in (b) the dopant atom is placed at the edge of the quantum dot.}\label{Fig:CEDiff}}
\end{figure}

Figure \ref{Fig:CE} shows the plot of the cohesive energy with quantum dot size for the different doping types and the two cases of dopant locations: a) center doping and b) edge doping. From these plots, we observe that the cohesive energy decreases with increasing quantum dot size. Pristine quantum dots have lower cohesive energy than doped clusters and are hence more stable. However, the cohesive energy difference between the doped and pristine clusters decreases with the increase in quantum dot size. This decrease in the difference in cohesive energy is primarily due to the decrease in dopant concentration with increasing nanocluster size{,} as shown in Figure \ref{Fig:CEDiff}. For a given quantum dot size, the cohesive energy of Al-doped > As-doped > P-doped > B-doped > pristine cluster. 

\begin{figure}[H]
\subfloat[Si$_{5}$H$_{12}$]{\includegraphics[keepaspectratio=true,width=0.35\textwidth]{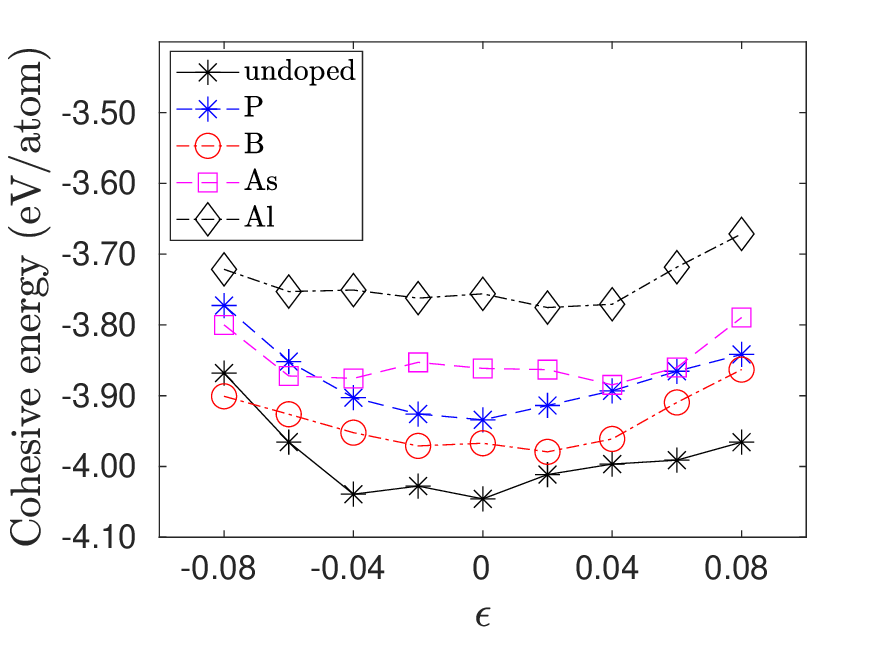}\label{Fig:S:CE1}}
\subfloat[Si$_{29}$H$_{36}$]{\includegraphics[keepaspectratio=true,width=0.35\textwidth]{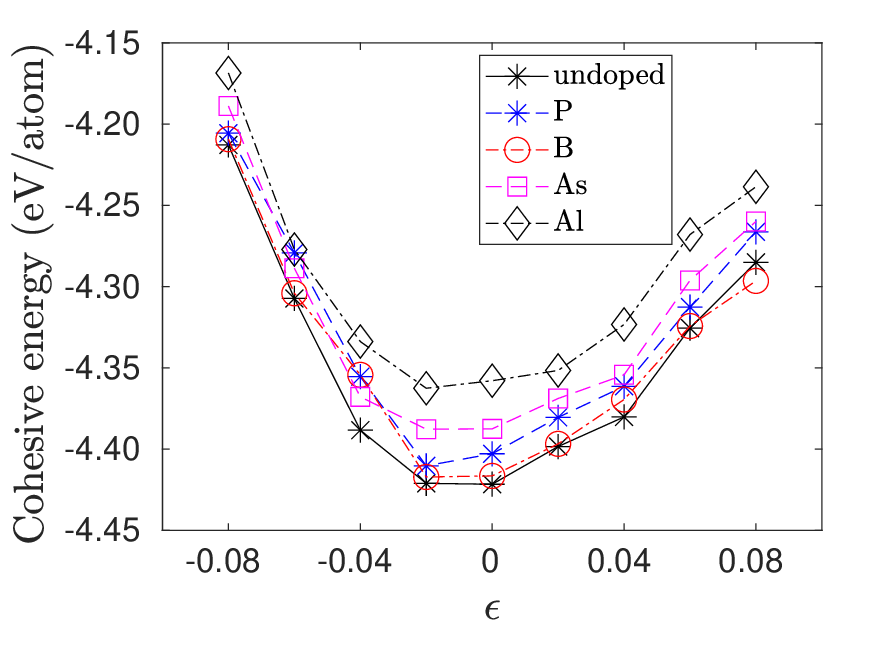}\label{Fig:S:CE2}}
\subfloat[Si$_{525}$H$_{276}$]{\includegraphics[keepaspectratio=true,width=0.35\textwidth]{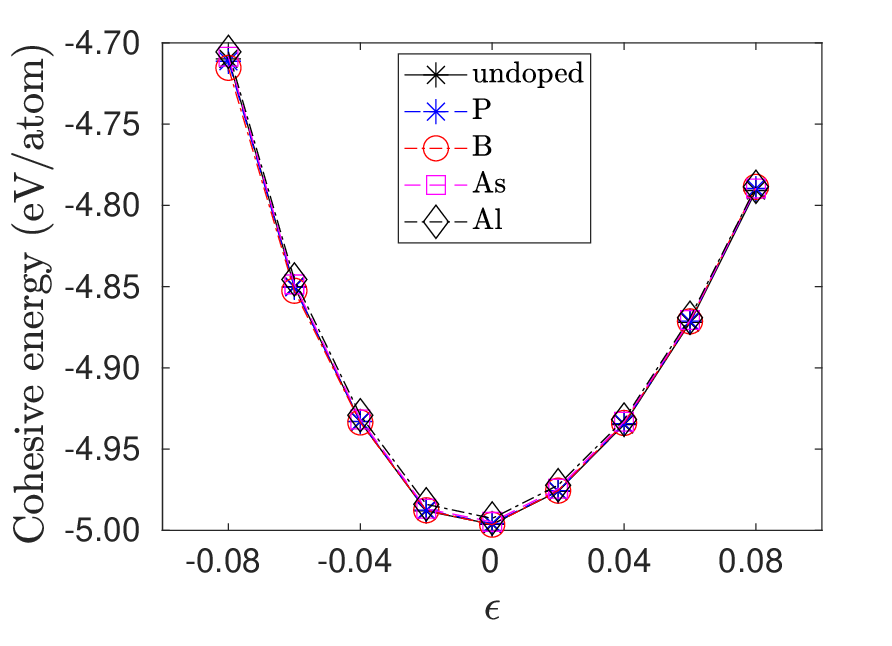}\label{Fig:S:CE3}} \\
\subfloat[Si$_{5}$H$_{12}$]{\includegraphics[keepaspectratio=true,width=0.35\textwidth]{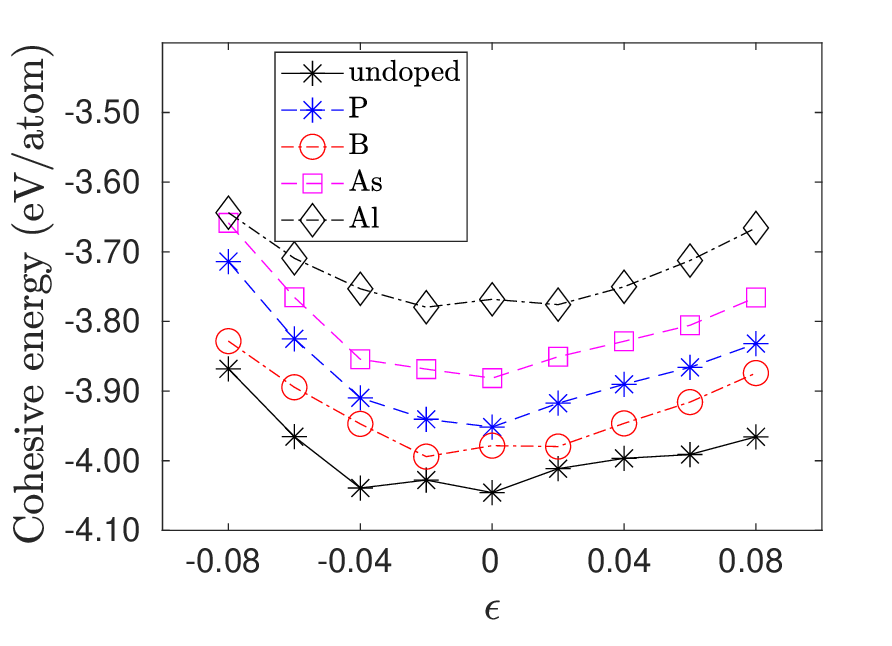}\label{Fig:S:CE4}}
\subfloat[Si$_{29}$H$_{36}$]{\includegraphics[keepaspectratio=true,width=0.35\textwidth]{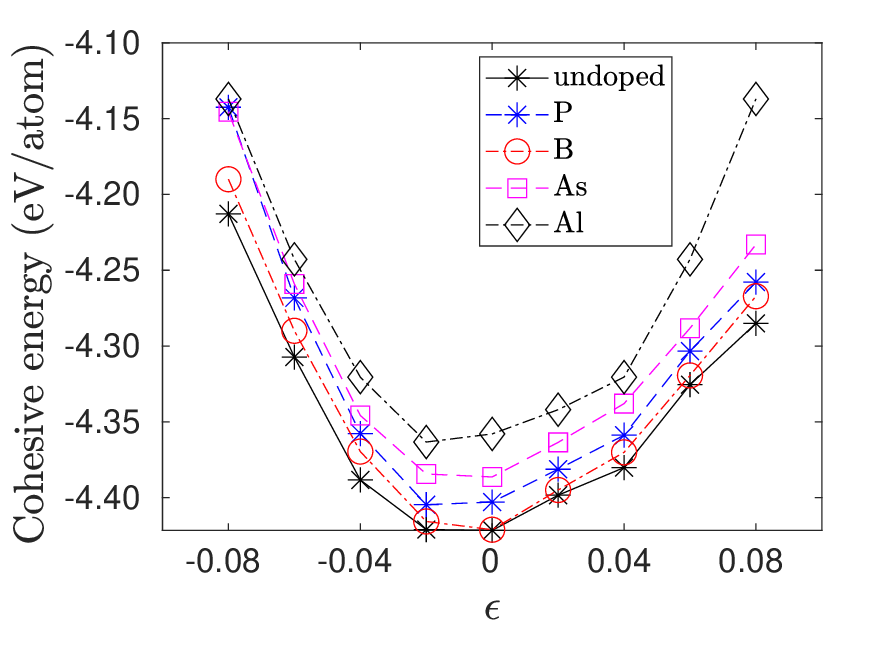}\label{Fig:S:CE5}}
\subfloat[Si$_{525}$H$_{276}$]{\includegraphics[keepaspectratio=true,width=0.35\textwidth]{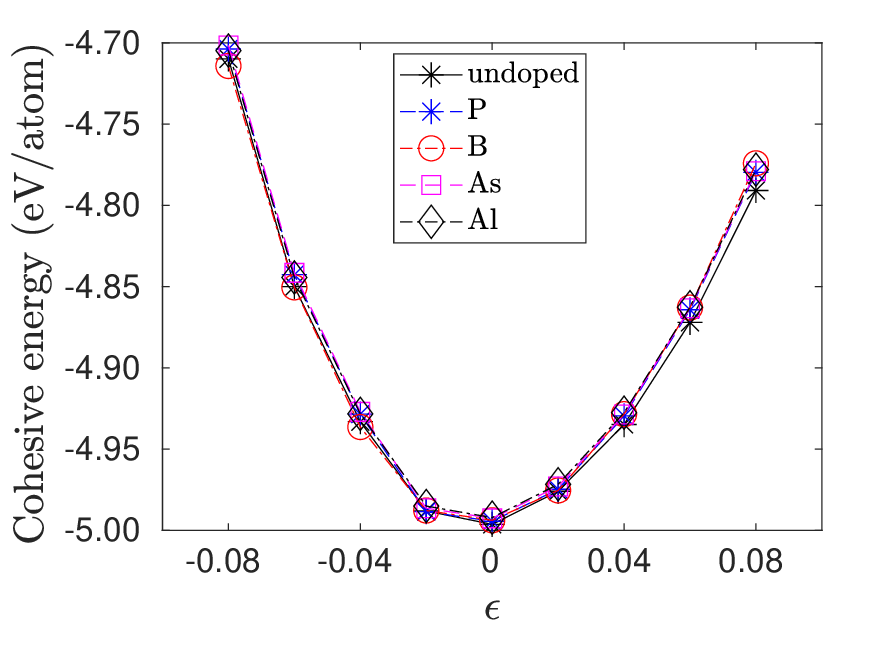}\label{Fig:S:CE6}}
{\caption{Varation of Cohesive energy of doped silicon quantum dots with strain $\epsilon$ for the smallest size (Si$_{5}$H$_{12}$), an intermediate size (Si$_{29}$H$_{36}$) and the largest size (Si$_{525}$H$_{276}$) considered in our study. In (a) - (c) the dopant atoms are placed at the center. In (d)-(f) the dopant atoms are placed at the edge.}\label{Fig:S:CE}}
\end{figure}

Figure \ref{Fig:S:CE} shows the effect of strain on the cohesive energy of doped and pristine silicon quantum dots. Both tensile and compressive strains increase the cohesive energy, implying a decrease in stability with strain. For the smallest quantum dot, the rate of increase in cohesive energy is symmetric with the direction of strain. However, as the size of the quantum dot increases, the quantum dots show an asymmetric response to tensile and compressive strains. The increase in cohesive energy is higher in compression than in tension with the same magnitude, which is similar to the response of bulk silicon under hydrostatic strain. Overall, the variation in cohesive energy with quantum dot size and strain can be expressed by
\begin{equation}
E_{coh}(d,\epsilon) = -d^n(c_0 + c_2 \epsilon^2 + c_4 \epsilon^4 )\,\,,
\end{equation}
where $n$, $c_0$, $c_2$ and $c_4$ are constants of the fit are shown in Table \ref{CohFit}, {and the fit is valid for strains $-8 \% \leq \epsilon \leq 8 \% $. } From this table, we also see that the cohesive energy of the edge-doped quantum dots is lower than the cohesive energy of the center-doped quantum dots, which is in agreement with the previous observations of edge-doped silicon quantum dots being more stable than center-doped ones \cite{pi:2012}.   

\begin{table}
\begin{center}
\caption{Constants of the fit of cohesive energy as a function of strain and quantum dot size. {$R^2$ and root mean square error (RMSE) of the fits are also presented.}}
\label{CohFit}
{\begin{tabular}{ c c c c c c c c}
\hline
\hline
dopant & location & $c_0$ & $c_2$ & $c_4$ & $n$ & $R^2$ & RMSE \\ 
\hline
\hline
undoped & - & $4.335$ & $-28.07$ & $-223.9$ & $0.1334$ & $0.9919$ & $0.02897$ \\
\hline
\multirow{2}{*}{P (n-type)}  & center & $4.279$ & $-30.14$ & $129.4$ & $0.1498$ & $0.9868$ & $0.04107$ \\
                    & edge & $4.276$ & $-34.06$& $87.99$ & $0.1502$ & $0.9868$ & $0.04107$\\
\hline
\multirow{2}{*}{B (p-type)}  & center & $4.307$& $-29.44$& $152.3$ & $0.1411$ & $0.9897$ & $0.03425$ \\
                    & edge & $4.306$ & $-31.43$ & $68.33$ & $0.1418$ & $0.9905$ & $0.03306$ \\
\hline
\multirow{2}{*}{As (n-type)}  & center & $4.256$ & $-22.91$ & $-760.4$ & $0.1538$ & $0.9837$ & $0.04664$ \\
                    & edge & $4.242$ & $-30.77$ & $-216.9$ & $0.1596$ & $0.9826$ & $0.05017$\\
\hline
\multirow{2}{*}{Al (p-type)}  & center & $4.200$ & $-26.01$ & $-263.9$ & $0.1700$ & $0.9758$ & $0.06257$ \\
                    & edge & $4.193$ & $-28.69$ & $-307.1$ & $0.1727$ & $0.9800$ & $0.05772$ \\
\hline
\hline
\end{tabular}}
\end{center}
\end{table}

\subsection{Energy gap}\label{Sec:BG}
\begin{figure}[H]
\subfloat[center]{\includegraphics[keepaspectratio=true,width=0.5\textwidth]{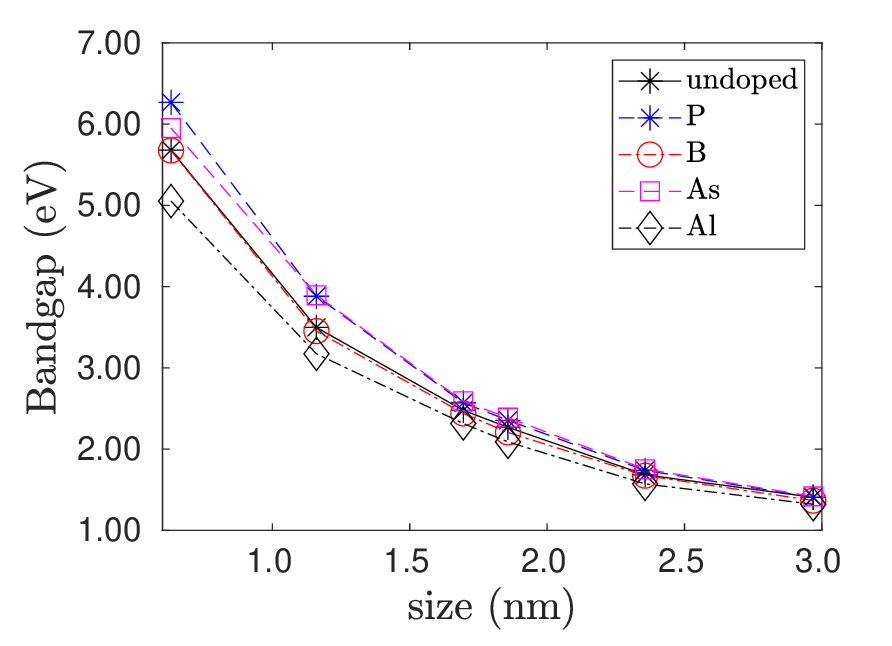}\label{Fig:R:BG}}
\subfloat[edge]{\includegraphics[keepaspectratio=true,width=0.5\textwidth]{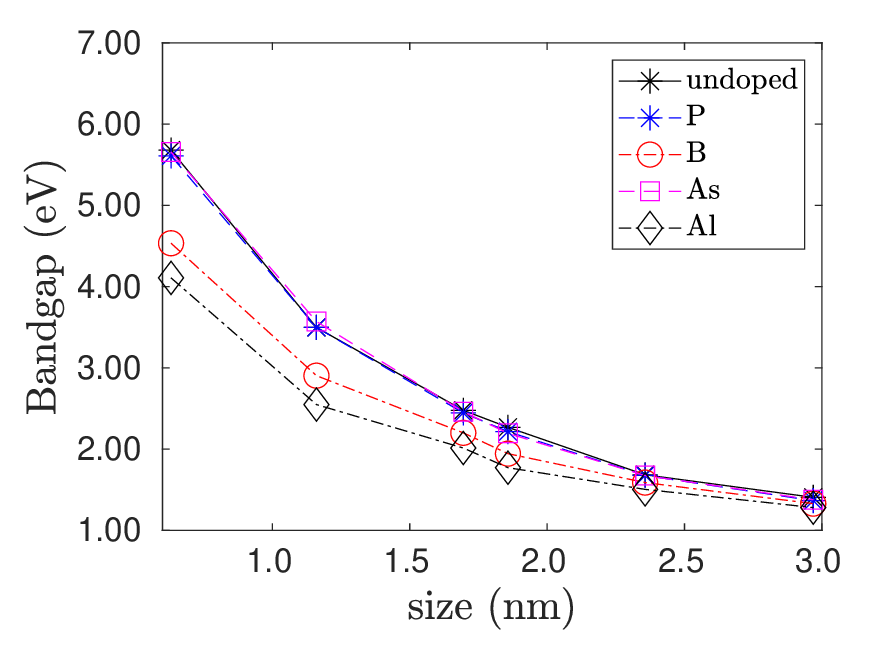}\label{Fig:R:BG:Edge}} \\
{\caption{Varation of energy gap of doped silicon quantum dots with quantum dot size. No strain is applied to the quantum dots. a) dopant {is} placed at the center of the quantum dot. b) dopant placed at the edge of the quantum dot.}\label{Fig:BG}}
\end{figure} 

\begin{figure}[H]
\subfloat[center]{\includegraphics[keepaspectratio=true,width=0.5\textwidth]{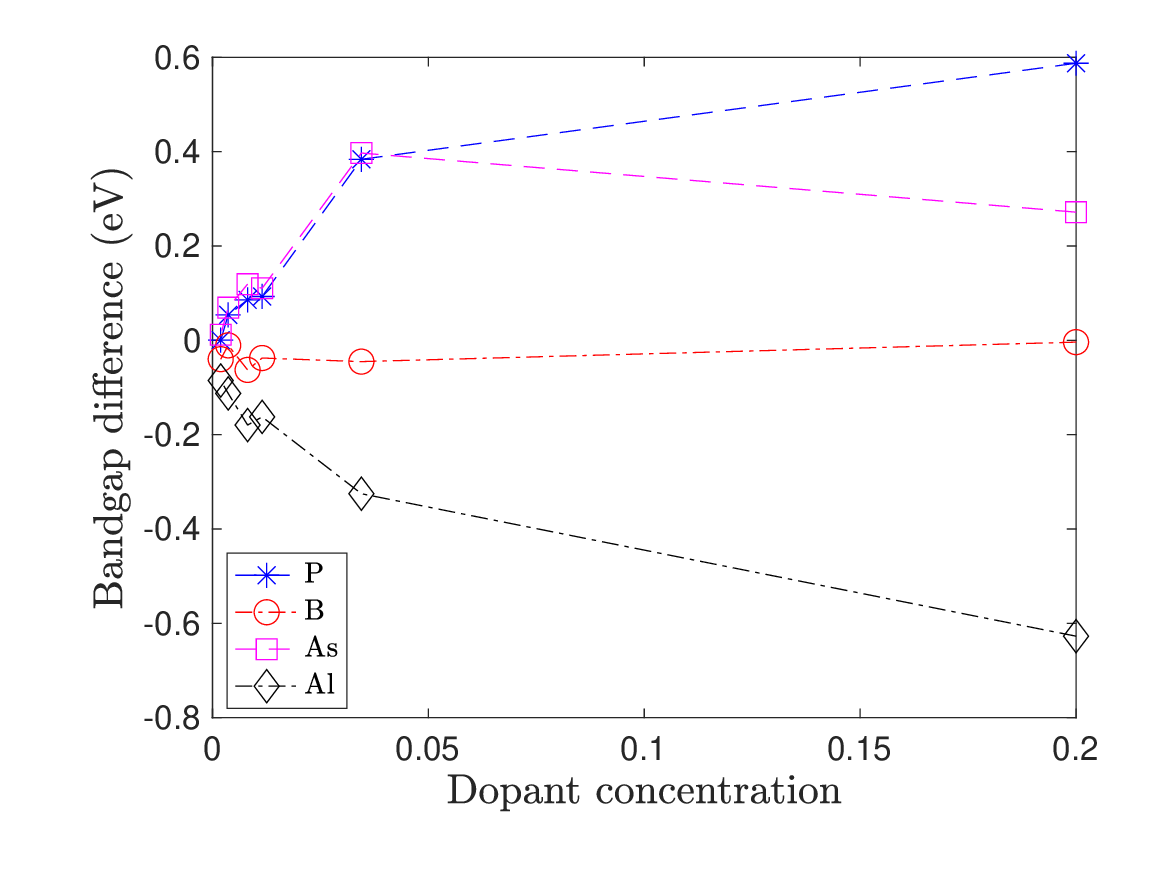}\label{Fig:C:BG}}
\subfloat[edge]{\includegraphics[keepaspectratio=true,width=0.5\textwidth]{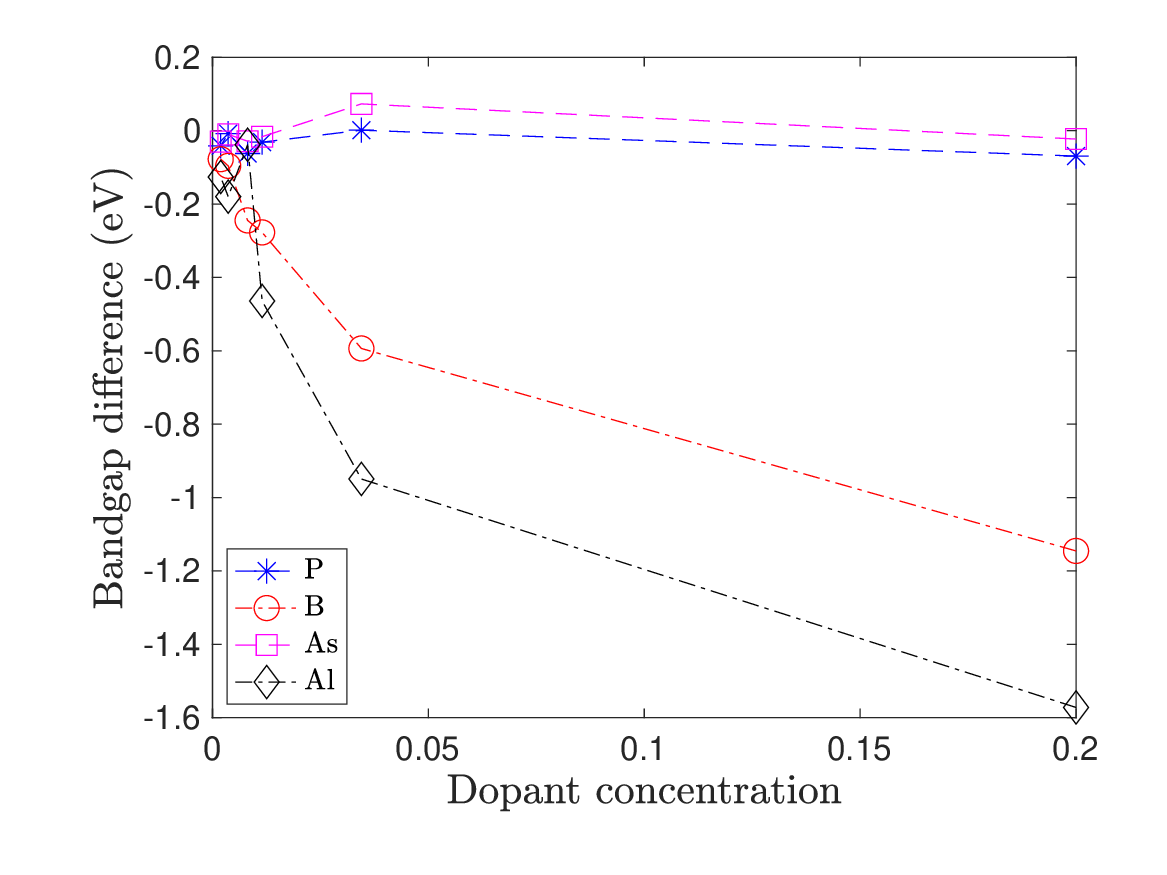}\label{Fig:C:BG:Edge}} \\
{\caption{Varation of the difference in energy gap of doped silicon quantum dots with dopant concentration. The difference is calculated with respect to the undoped case. No strain is applied to the quantum dots. a) dopant {is} placed at the center of the quantum dot. b) dopant placed at the edge of the quantum dot.}\label{Fig:BGDiff}}
\end{figure} 

Figure \ref{Fig:BG} shows the plot of the variation of energy gap -- sometimes also referred to as bandgap -- with quantum dot size for doped and pristine silicon clusters. The energy gap is defined as the difference between the highest occupied molecular orbital (HOMO) and lowest unoccupied molecular orbital (LUMO) energies. The Kohn-Sham HOMO-LUMO orbital energy difference approximates the excitation energy difference of a molecule and is a good measure of its optical gap \cite{baerends:2013}. The energy gaps for the pristine case are in good agreement with previous results \cite{peng:2006}. In general, the energy gap decreases with quantum dot size. The effect of doping not only depends on the type and location of the dopant but also on the quantum dot size. From Figure \ref{Fig:R:BG}, when the dopant is placed at the center, we observe that, for quantum dot sizes of less than 2 nm, n-type doping with P and As increases the energy gap, and p-type doping with Al decreases the energy gap. Furthermore, p-type doping with B has {a} negligible effect on the energy gap. Figure \ref{Fig:R:BG:Edge} shows the variation of energy gap when the dopant is placed at the edge. For quantum dot sizes less than 2 nm, p-type dopants (B and Al) decrease the energy gaps, whereas n-type dopants (P and As) do not significantly affect the energy gaps. Overall, the effect of the dopants on the energy gap decreases with an increase in quantum dot size. This decrease in the difference in energy gap is due to the reduction of dopant concentration with {an} increase in nanocluster size as shown in Figure \ref{Fig:BGDiff}.

\begin{figure}[H]
\subfloat[Si$_{5}$H$_{12}$]{\includegraphics[keepaspectratio=true,width=0.35\textwidth]{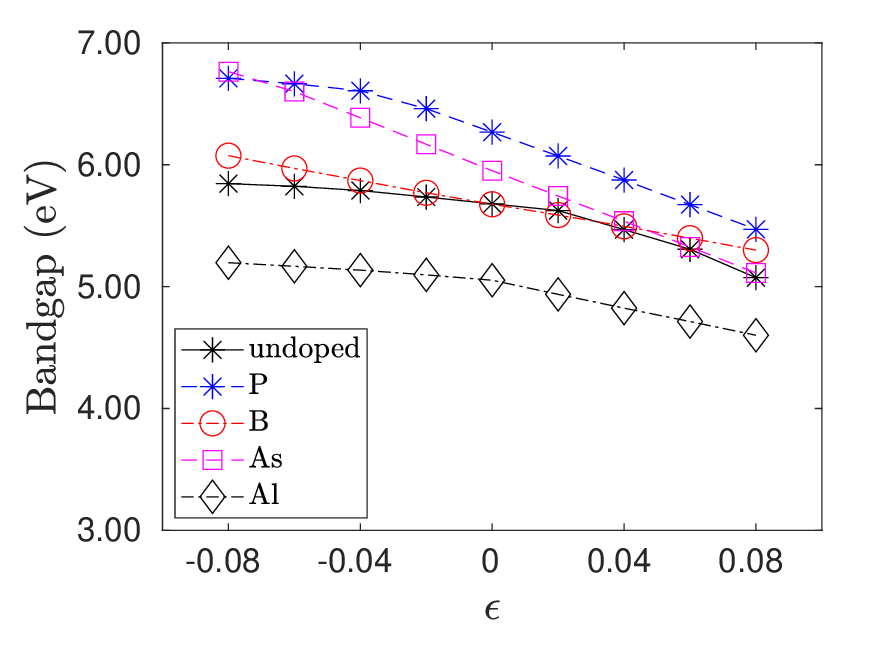}\label{Fig:S:BGa}}
\subfloat[Si$_{29}$H$_{36}$]{\includegraphics[keepaspectratio=true,width=0.35\textwidth]{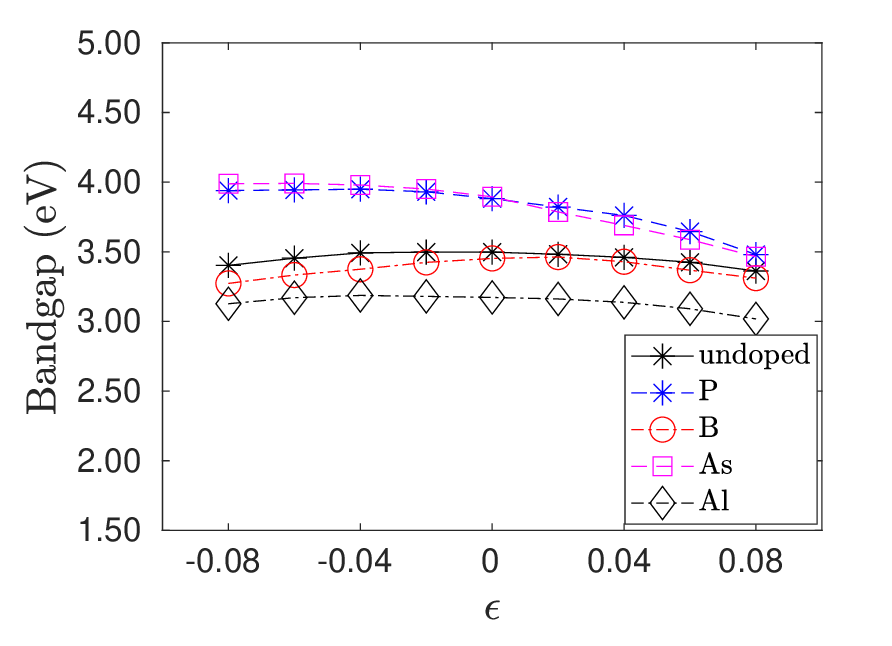}\label{Fig:S:BGb}}
\subfloat[Si$_{525}$H$_{276}$]{\includegraphics[keepaspectratio=true,width=0.35\textwidth]{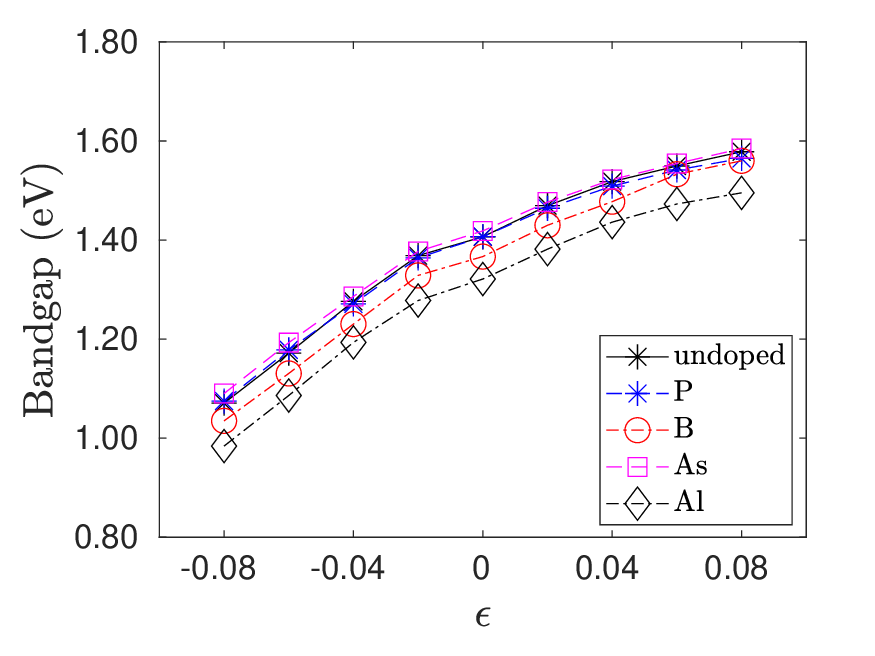}\label{Fig:S:BGc}} \\
\subfloat[Si$_{5}$H$_{12}$]{\includegraphics[keepaspectratio=true,width=0.35\textwidth]{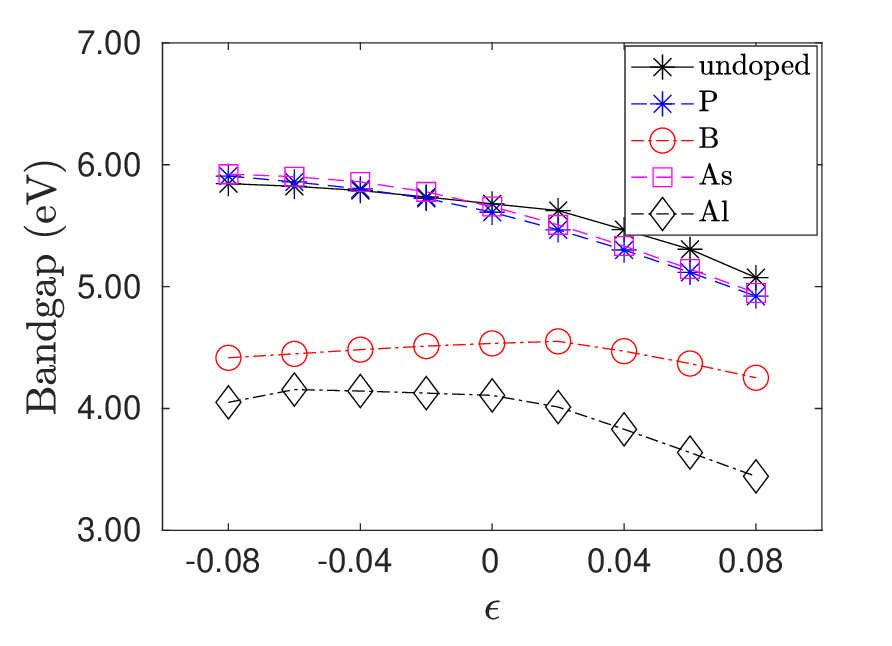}\label{Fig:S:BGd}}
\subfloat[Si$_{29}$H$_{36}$]{\includegraphics[keepaspectratio=true,width=0.35\textwidth]{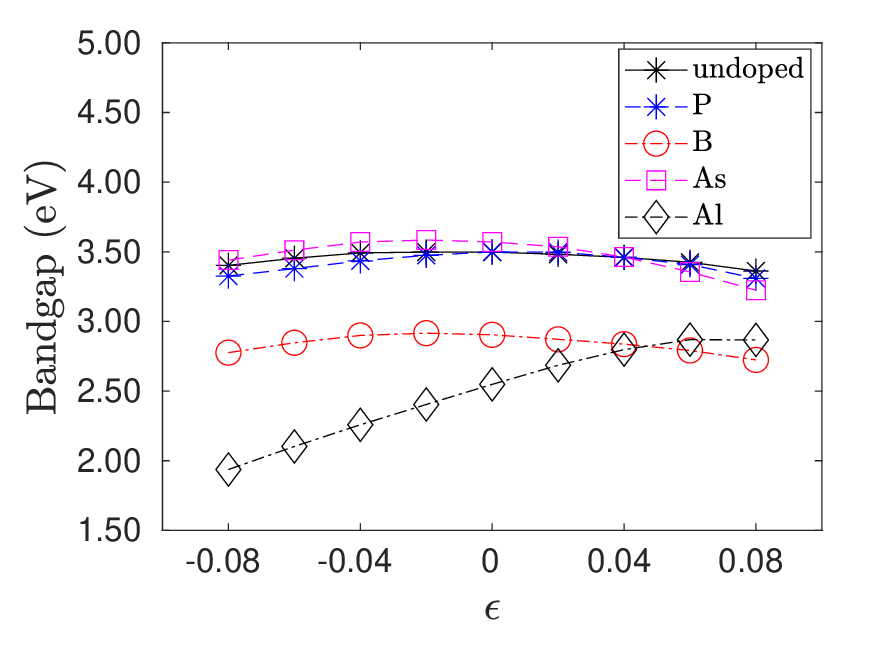}\label{Fig:S:BGe}}
\subfloat[Si$_{525}$H$_{276}$]{\includegraphics[keepaspectratio=true,width=0.35\textwidth]{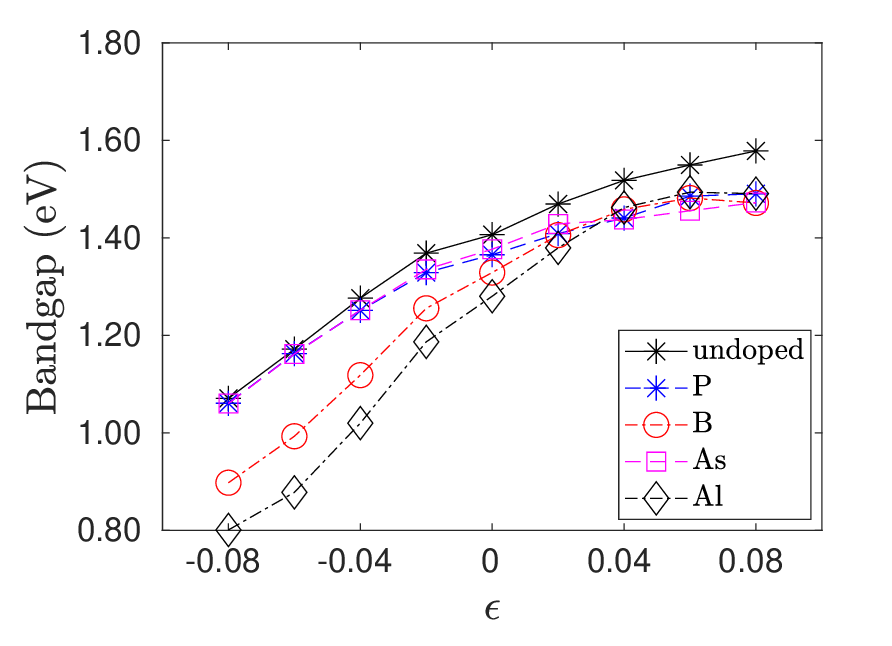}\label{Fig:S:BGf}}
{\caption{Varation of energy gap of doped silicon quantum dots with strain $\epsilon$ for the smallest size (Si$_{5}$H$_{12}$), intermediate size (Si$_{29}$H$_{36}$) and the largest size (Si$_{525}$H$_{276}$) considered in our study.  In (a) - (c), the dopant atoms are placed at the center. In (d)-(f), the dopant atoms are placed at the edge.}\label{Fig:S:BG}}
\end{figure}

Figure \ref{Fig:S:BG} shows the variation of energy gap with strain. The nature of the effect of strain on the energy gap of silicon quantum dots depends on the size of the quantum dot. For the smallest quantum dot, the energy gap decreases with increasing tensile strain and increases with compressive strain. For quantum dots of size greater than 2 nm, the energy gap increases with tension and decreases with compression. Quantum dots of sizes ranging from 1-2 nm show intermediate behavior. The effect of strain on the energy gaps of pristine silicon quantum dots is in excellent agreement with previous DFT studies \cite{peng:2006}. Figure \ref{Fig:S:BG} also shows the variation in energy gap for the doped quantum dots. Except for Al-doping {(edge)} in the 1-2 nm size range, doped quantum dots show similar behavior as pristine quantum dots. As seen from Figure \ref{Fig:S:BGe}, in Al-doped quantum dots of intermediate sizes 1-2 nm, the energy gap decreases with compressive strain and increases with tensile strain. It is also observed that the energy gap is smaller in the edge-doped quantum dots compared to the center-doped ones.

{The asymmetric response of quantum dots to strain can be understood by the changes in HOMO and LUMO energies \cite{peng:2006}. DFT calculations show that for large quantum dots, the LUMO energy increases with compressive strain and decreases with tension but the HOMO energy is unaffected. This results in an increase (decrease) in the energy gap with tension (compression). Similar trend in HOMO and LUMO is also observed for intermediate sized Al-doped (edge) quantum dots. For small quantum dots, the LUMO energy is unaffected by strain, but the HOMO energy decreases with tension and increases with compression, resulting in an overall increase (decrease) in the energy gap in compression (tension). For the intermediate sized quantum dots, HOMO and LUMO energies are similarly affected by tensile and compressive strains.}
\begin{figure}[H]\centering
\subfloat[HOMO orbitals]{\includegraphics[keepaspectratio=true,width=0.7\textwidth]{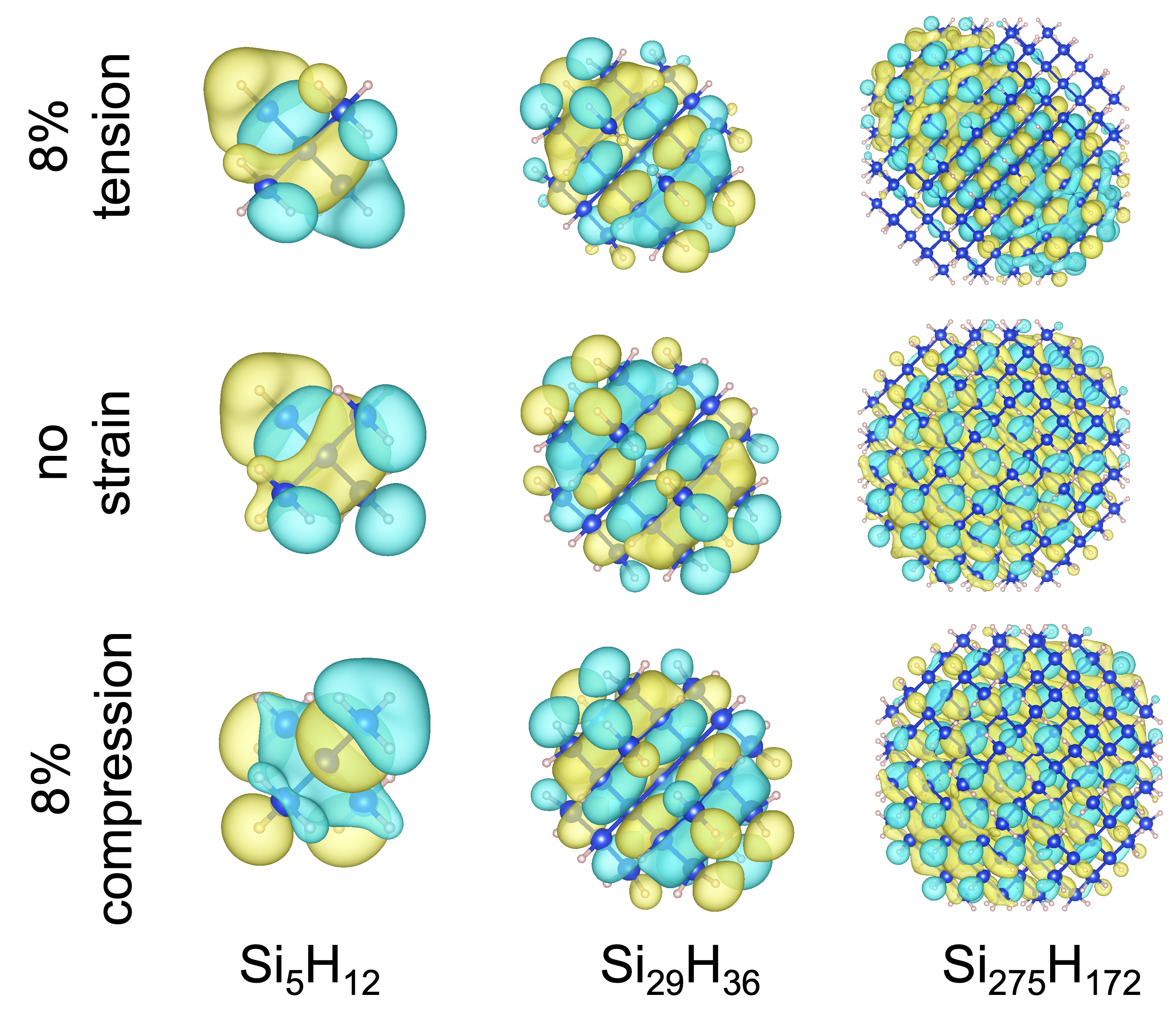}}\\
\subfloat[LUMO orbitals]{\includegraphics[keepaspectratio=true,width=0.7\textwidth]{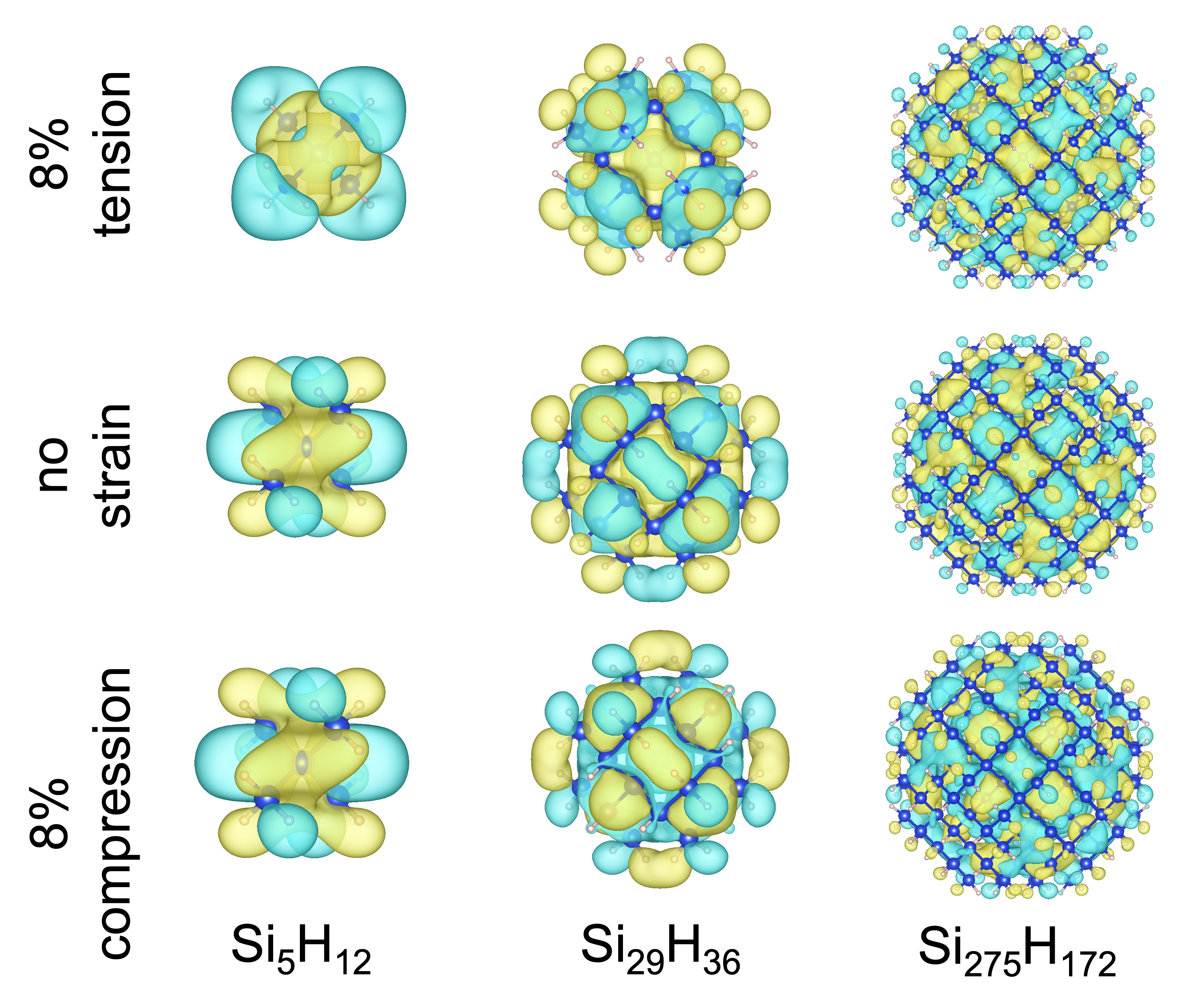}}
\caption{(a)HOMO and (b) LUMO orbitals of small, intermediate and large sized silicon quantum dot under no strain and 8\% tensile and compressive strain. Yellow and green colors denote positive and negative values, respectively.} \label{Fig:HOMOLUMO}
\end{figure}

In Figure \ref{Fig:HOMOLUMO}, we show the effect of strain on the HOMO and LUMO orbitals for small, intermediate and large quantum dots, plotted using the VESTA \cite{VESTA} software package. {An orbital shows bonding character when it is centered between two atoms, and shows anti-bonding character when it is centered on the atoms.} From these figures, it is clear that in small quantum dots, HOMO shows bonding character and LUMO shows anti-bonding character. As compressive strains decrease the Si-Si bond lengths whereas tensile strains increase the Si-Si bond lengths, HOMO shows more bonding character in compression and some anti-bonding character in tension. The effect of strain on the anti-bonding character of LUMO for small quantum dots is negligible{,} as the electrons are already localized in the vicinity of the atoms. LUMO orbitals show a mixture of bonding and anti-bonding character for intermediate-sized quantum dots. Tensile strains reduce the bonding character of LUMO, while compressive strains slightly increase the bonding character of LUMO. The HOMO orbitals for intermediate sized quantum dots show bonding character, which is enhanced in compression and depleted in tension. For the large sized quantum dots, both HOMO and LUMO orbitals show a mixture of bonding and anti-bonding characters. Under compressive (tensile) strains, the bonding character of these orbitals is increased (decreased).

Overall, the variation in energy gap with quantum dot size $d$ and strain $\epsilon$ can be expressed using the form 
\begin{equation}\label{Egfit}
E_g(d,\epsilon) = \frac{A(\epsilon)}{d^n}+B(\epsilon) \,\,,
\end{equation}
where $A(\epsilon)$ and $B(\epsilon)$ are polynomial functions of the form
\begin{eqnarray}
A(\epsilon) &=& \frac{a_3}{3} \epsilon^3 + \frac{a_2}{2} \epsilon^2 + a_0 \,\,, \\
B(\epsilon) &=& \frac{b_3}{3} \epsilon^3 + b_0 \,\,.
\end{eqnarray}
where $b_0=0.45$ eV is the LDA energy gap of Silicon \cite{nogueira2003tutorial} and the constants $a_0$, $a_2$, $a_3$, and $b_3$ and the power $n$ are presented in Table \ref{GapFit}. The fit to the bandgap is valid for strains $-8 \% \leq \epsilon \leq 8 \% $. Notice that the energy gap is almost inversely related to the diameter of the quantum dot. The power $n=1.0138$ calculated in this work is in excellent agreement with previously calculated values of $n=1.0$ \cite{delley:1995} and $n=1.1$ \cite{ougut:1997}. To the author's knowledge, scaling laws for doped and strained cases have not been reported before this work. We validate the fit in Equation \ref{Egfit} by calculating the energy gap of a 1 nm diameter quantum dot as 3.7827 eV which is in good agreement with the experimentally reported value of 3.5 eV \cite{smith:2005} and 2 nm diameter as 2.1005 eV being in excellent agreement with the experimental value of 2.1 eV \cite{wilcoxon:1999}.

\begin{table}
\caption{Constants of the fit of energy gap as a function of strain and quantum dot size. {$R^2$ and root mean square error (RMSE) of the fits are also presented.}}
\label{GapFit}
\begin{center}
{\begin{tabular}{c c c c c c c c c c}
\hline
\hline
dopant & location &$a_0$ & $a_2$ & $a_3$ &  $b_3$ & $n$ & $R^2$ & RMSE \\ 
       &          &   & & $\times 10^3$ & $\times 10^3$  & \\
\hline
\hline
undoped & - & $3.3327$ &$-47.4469$& $-3.2671$ & $2.6202$ & $1.0138$ & $0.9906$ & $0.1443$ \\
\hline 
\multirow{2}{*}{P (n-type)}  & center & $3.6331$ & $-46.0618$ & $-4.5675$ & $2.9590$ & $1.0706$ & $0.9858$ & $0.2255$ \\
                    & edge & $3.2543$ & $-50.6553$ & $-3.5408$ & $2.5376$ & $1.0426$ & $0.9885$ & $0.1761$ \\
 \hline
 \multirow{2}{*}{B (p-type)}  & center & $3.2837$ & $-20.4403$ & $-3.2731$ & $2.7176$ & $1.0528$ & $0.9928$ & $0.1301$ \\
                    & edge & $2.7000$ & $-64.0062$ & $-2.2641$ & $2.6949$ & $0.9480$  & $0.9886$ & $0.1201$ \\
\hline
 \multirow{2}{*}{As (n-type)}  & center & $3.5502$ & $-25.6341$ & $-5.8658$ & $3.6007$ & $1.0175$ & $0.9805$ & $0.2286$ \\
                    & edge & $3.3178$ & $-65.5933$ & $-3.5655$ & $2.3284$ & $1.0321$  & $0.9865$ & $0.1750$ \\
 \hline
 \multirow{2}{*}{Al (p-type)}  & center & $2.9608$ & $-36.6833$ & $-2.9018$ & $2.3460$ & $0.9866$ & $0.9884$ & $0.1402$ \\
                    & edge & $2.4634$ & $-74.1392$ & $-3.9649$ & $4.6290$ & $0.8628$  & $0.9393$ & $0.2402$ \\
 \hline
\end{tabular}}
\end{center}
\end{table}

\subsection{Fermi energy change}\label{Sec:FE}
\begin{figure}[H]
\subfloat[center]{\includegraphics[keepaspectratio=true,width=0.5\textwidth]{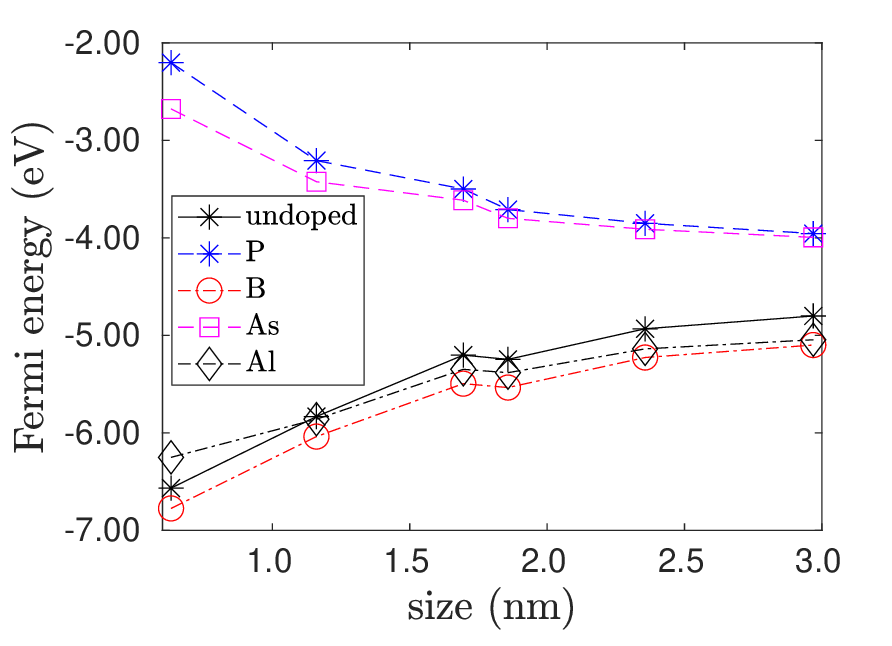}\label{Fig:R:FE}} 
\subfloat[edge]{\includegraphics[keepaspectratio=true,width=0.5\textwidth]{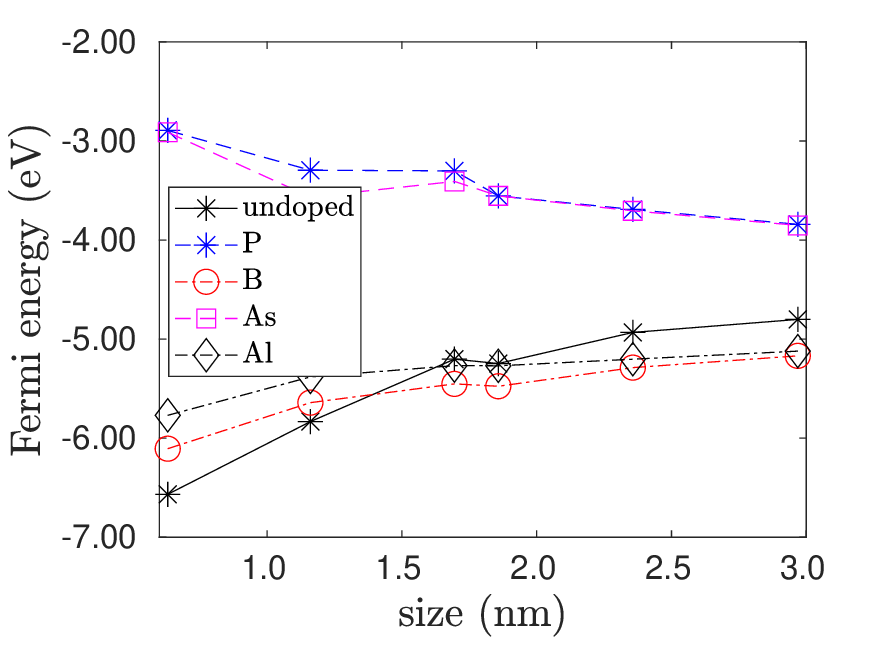}\label{Fig:R:FE:Edge}}\\
{\caption{{Variation} of Fermi energy of doped silicon quantum dots with quantum dot size. No strain is applied {to} the quantum dots. In (a) dopant {is} placed at the center of the quantum dot{,} and in (b) dopant {is} placed at the edge of the quantum dot.}\label{Fig:FERMI}}
\end{figure}

Next, we present the change in Fermi energy with strain and quantum dot size. Figure \ref{Fig:FERMI} shows the variation of Fermi energy with quantum dot size. For pristine quantum dots, the Fermi energy increases with quantum dot size. A similar trend is also observed for p-type doping with B and Al atoms. For n-type doping with P and As the Fermi energy decreases when the dopant atom is placed at the center.

\begin{figure}[H]
\subfloat[Si$_{5}$H$_{12}$]{\includegraphics[keepaspectratio=true,width=0.35\textwidth]{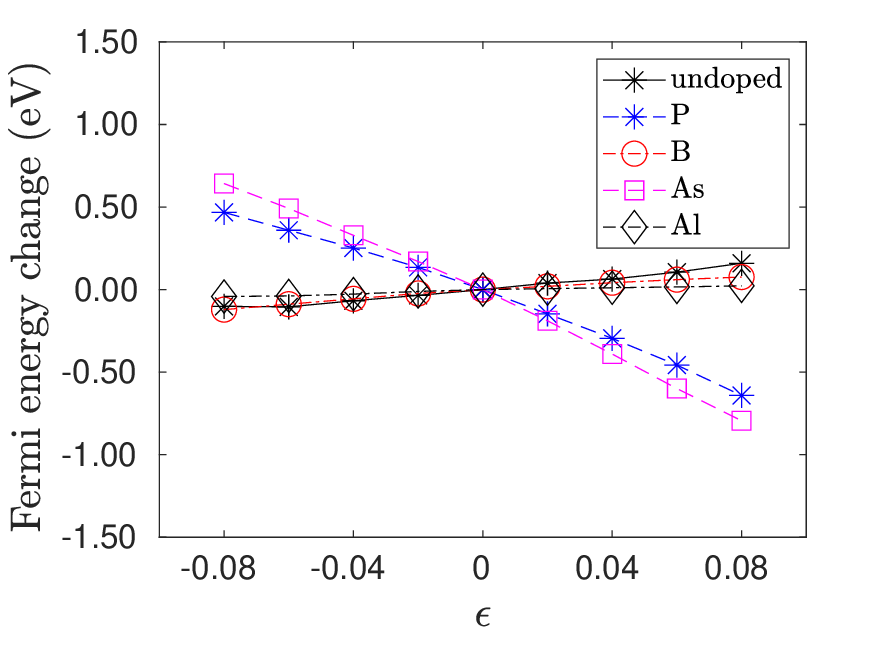}\label{Fig:S:FEa}}
\subfloat[Si$_{29}$H$_{36}$]{\includegraphics[keepaspectratio=true,width=0.35\textwidth]{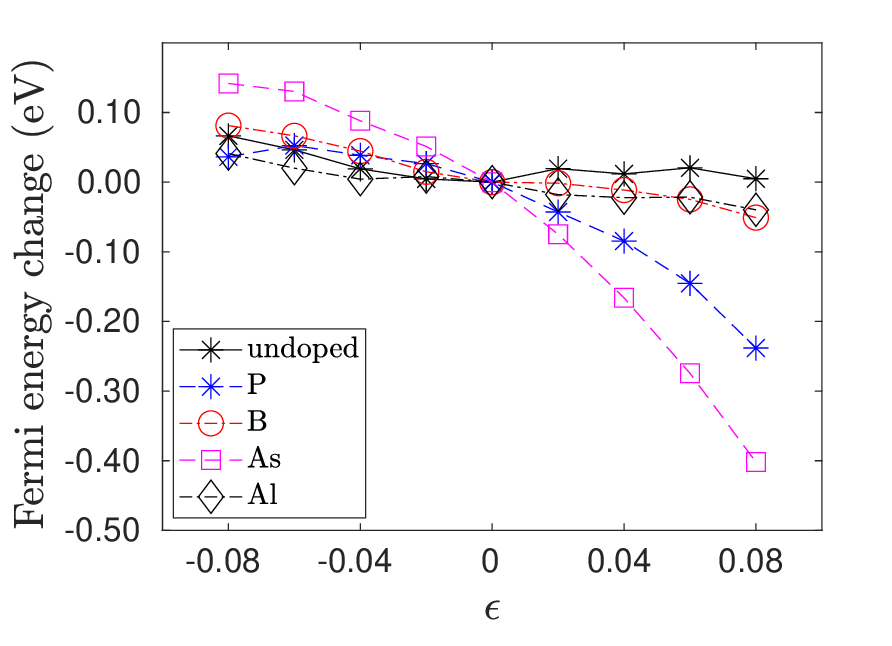}\label{Fig:S:FEb}}
\subfloat[Si$_{525}$H$_{276}$]{\includegraphics[keepaspectratio=true,width=0.35\textwidth]{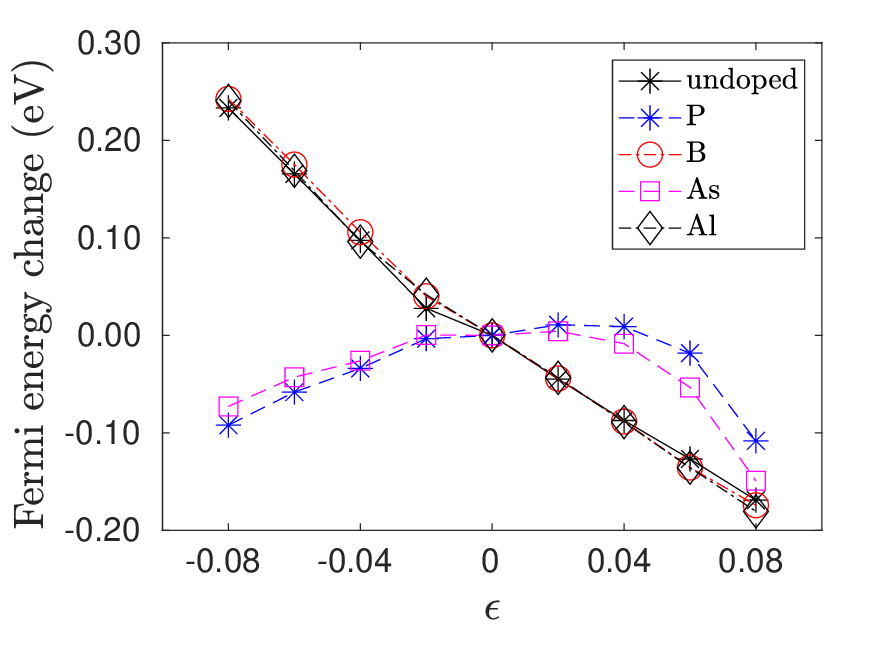}\label{Fig:S:FEc}} \\
\subfloat[Si$_{5}$H$_{12}$]{\includegraphics[keepaspectratio=true,width=0.35\textwidth]{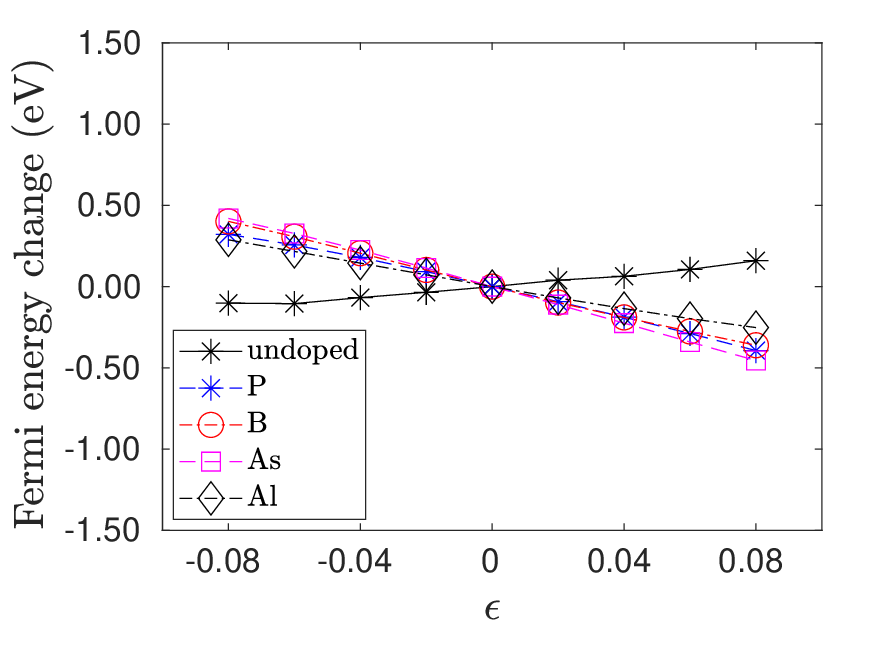}\label{Fig:S:FEd}}
\subfloat[Si$_{29}$H$_{36}$]{\includegraphics[keepaspectratio=true,width=0.35\textwidth]{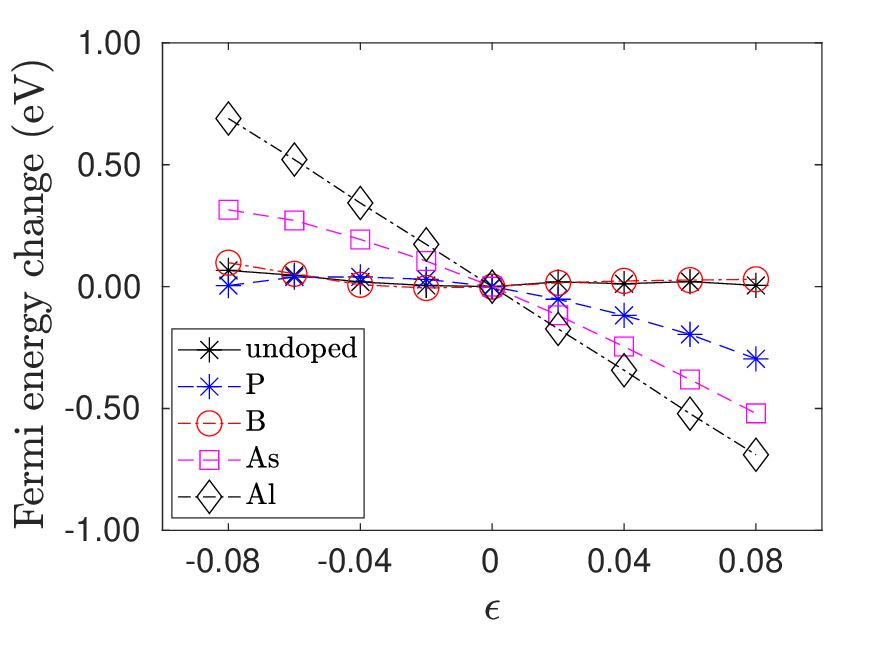}\label{Fig:S:FEe}}
\subfloat[Si$_{525}$H$_{276}$]{\includegraphics[keepaspectratio=true,width=0.35\textwidth]{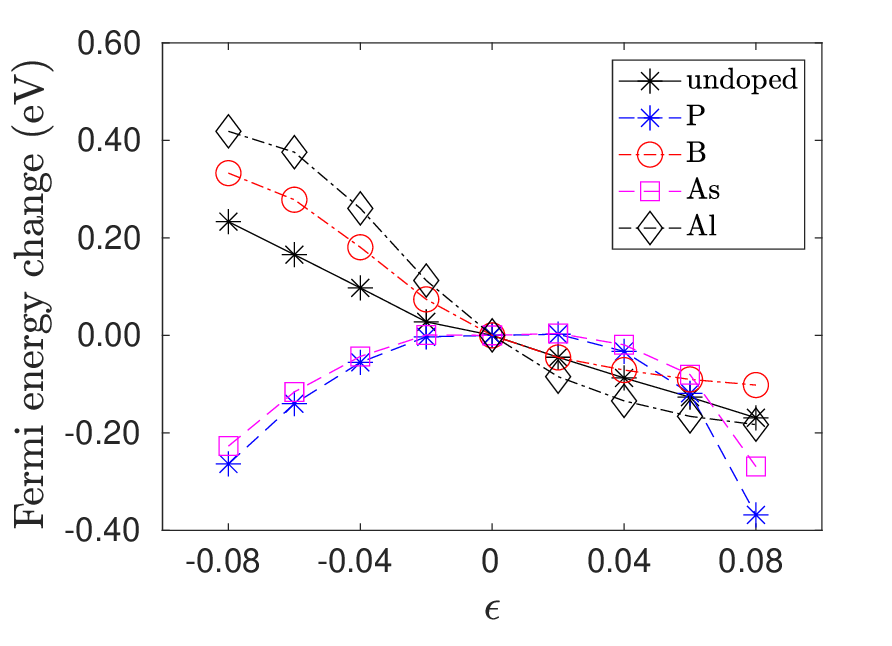}\label{Fig:S:FEf}}
{\caption{Change in Fermi energy of doped silicon quantum dots with strain $\epsilon$ for the smallest size (Si$_{5}$H$_{12}$), an intermediate size (Si$_{29}$H$_{36}$) and the largest size (Si$_{525}$H$_{276}$) considered in our study. In (a) - (c), the dopant atoms are placed at the center. In (d)-(f), the dopant atoms are placed at the edge.}\label{Fig:S:FE}}
\end{figure}
Figure \ref{Fig:S:FE} shows the change in Fermi levels with strain for doped and pristine silicon quantum dots of different sizes and dopant positions. We find that the response of the Fermi level to strain is not only dependent on the quantum dot size but also on the dopant type and position. We first consider the smallest quantum dot (Si$_5$H$_{12}$). Figure \ref{Fig:S:FEa} and \ref{Fig:S:FEd}  shows the cases when the dopant atoms are placed at the center and at the edge, respectively.  From Figure \ref{Fig:S:FEa}, we see that, for pristine and p-type doping with B and Al at the center, the Fermi level increases with tension and decreases with compression, whereas for n-type doping with P and As at center, tensile strains decrease and compressive strains increase the Fermi energy. From Figure \ref{Fig:S:FEd}, when a dopant is at the edge, both p- and n-type doped quantum dots show similar behavior, where tensile strains decrease and compressive strains increase the Fermi energy. 

Next, we consider the doped quantum dots of intermediate size 1-2 nm. Taking Si$_{29}$H$_{36}$ as a representative example, Figure \ref{Fig:S:FEb} and \ref{Fig:S:FEe}  shows the cases when the dopant atoms are placed at the center and at the edge, respectively. From these figures, we see that for pristine clusters, strains do not effect the Fermi level. However, for n-type doping at the center (Figure\ref{Fig:S:FEb}) and at the edge (Figure \ref{Fig:S:FEe}), tensile strain decreases the Fermi level{,} but compressive strains have {a} negligible effect. Intermediate-sized quantum dots doped at the center and edge with B atoms (p-type) do not show a change in Fermi energy with strain. However, when intermediate-sized quantum dots are doped with Al atoms (p-type), the response of the Fermi energy to strain depends on the location of the dopant. As such, from Figure \ref{Fig:S:FEb},  quantum dots with Al dopant at the center show a negligible change in Fermi energy with strain. However, from Figure \ref{Fig:S:FEe}{,} when {the} Al dopant is at the edge, tensile strains decrease the Fermi energy{,} and compressive strains increase the Fermi energy. 

Finally, we consider quantum dots with sizes greater than 2 nm. Taking Si$_{525}$H$_{276}$ as an example, we see from Figs. \ref{Fig:S:FEc} and \ref{Fig:S:FEf}, for pristine and p-type doped quantum dots, tensile strains decrease and compressive strains increase the Fermi energy{,} and for n-type doped quantum dots, both tensile and compressive strains lower the Fermi energy. 

\subsection{Density of states and electron density}\label{Sec:DOS}
\begin{figure}[H]\centering
\subfloat[Si$_{29}$H$_{36}$ pristine]{\includegraphics[keepaspectratio=true,width=0.45\textwidth]{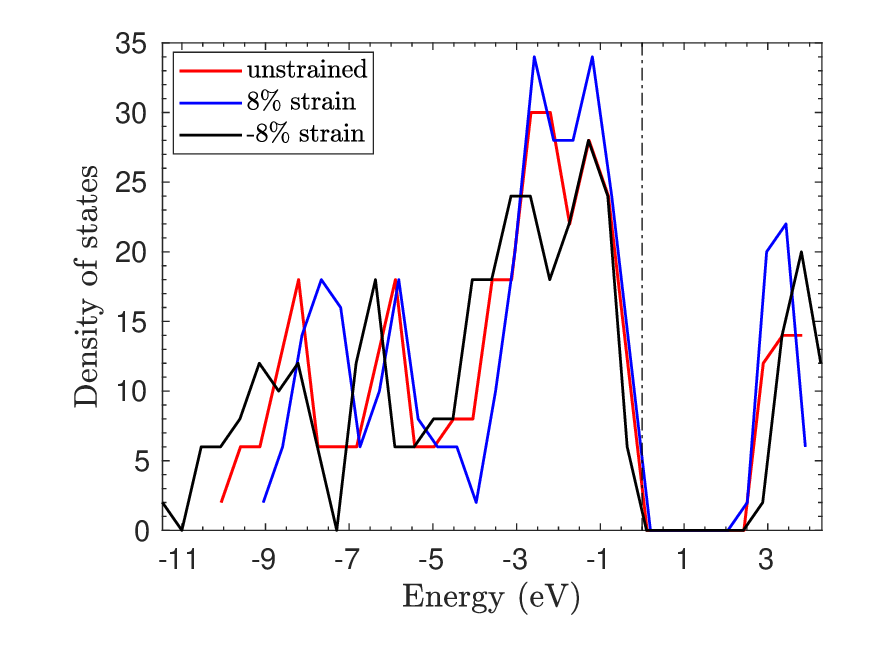}\label{Fig:DOSa}}
\subfloat[Si$_{525}$H$_{276}$ pristine]{\includegraphics[keepaspectratio=true,width=0.45\textwidth]{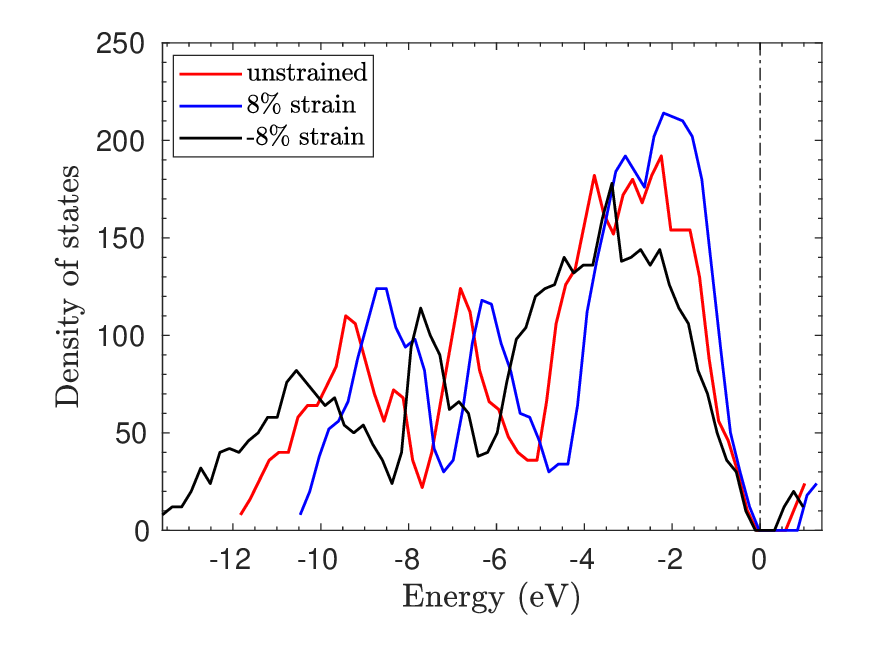}\label{Fig:DOSb}} \\
\subfloat[Si$_{29}$H$_{36}$ P doped (n- type)]{\includegraphics[keepaspectratio=true,width=0.45\textwidth]{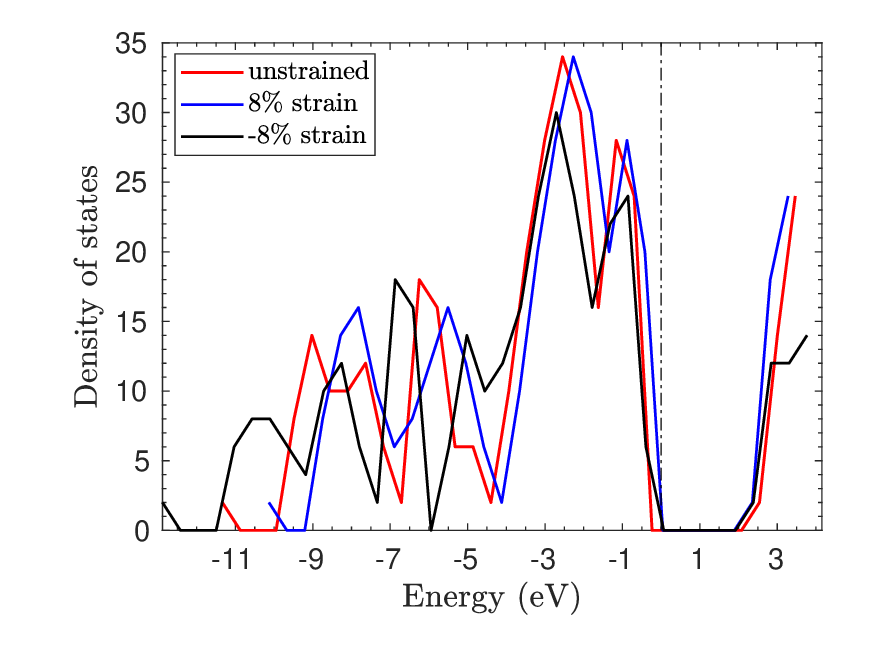}\label{Fig:DOSc}}
\subfloat[Si$_{525}$H$_{276}$ P doped (n- type)]{\includegraphics[keepaspectratio=true,width=0.45\textwidth]{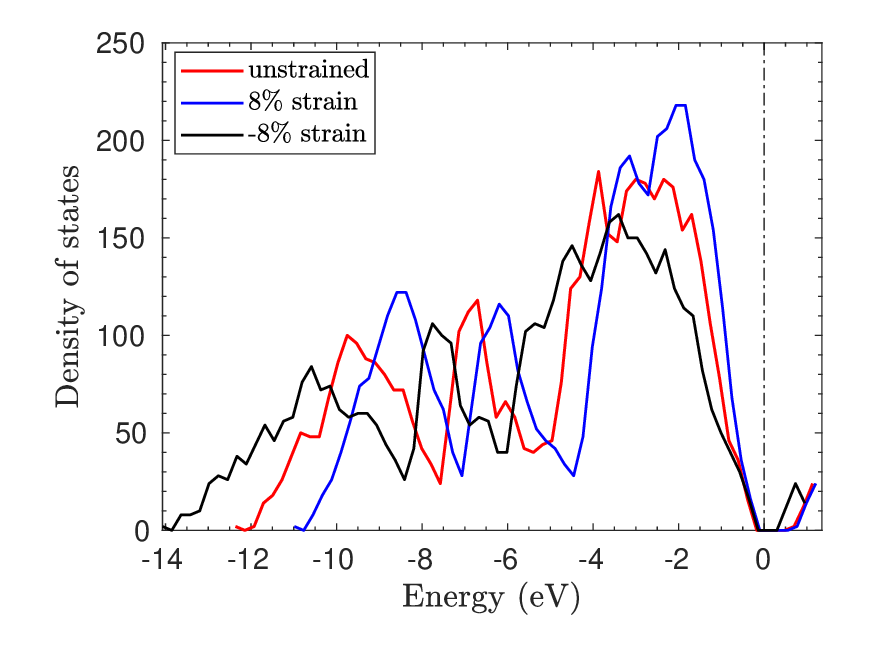}\label{Fig:DOSd}} \\
\subfloat[Si$_{29}$H$_{36}$ B doped (p- type)]{\includegraphics[keepaspectratio=true,width=0.45\textwidth]{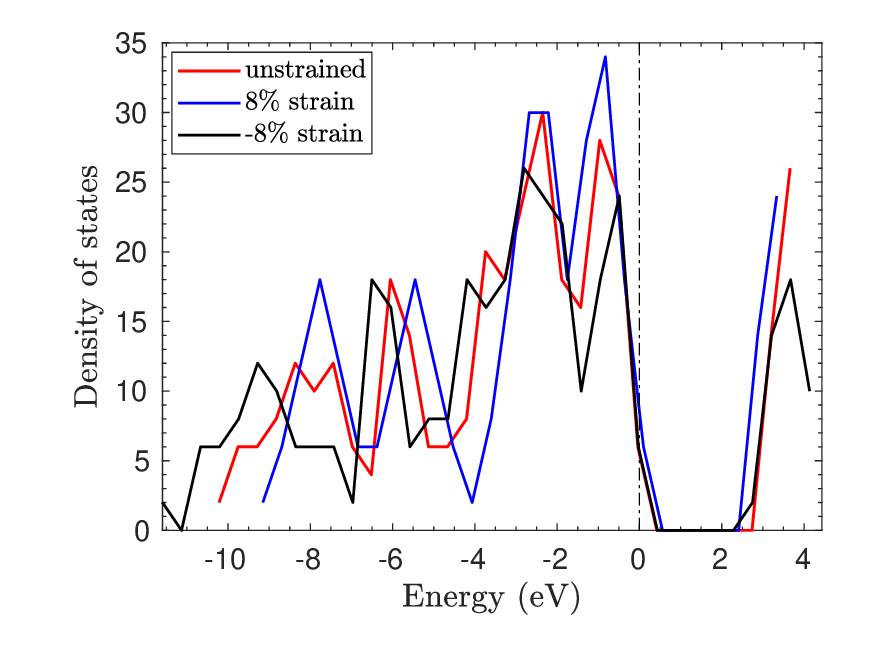}\label{Fig:DOSe}}
\subfloat[Si$_{525}$H$_{276}$ B doped (p- type)]{\includegraphics[keepaspectratio=true,width=0.45\textwidth]{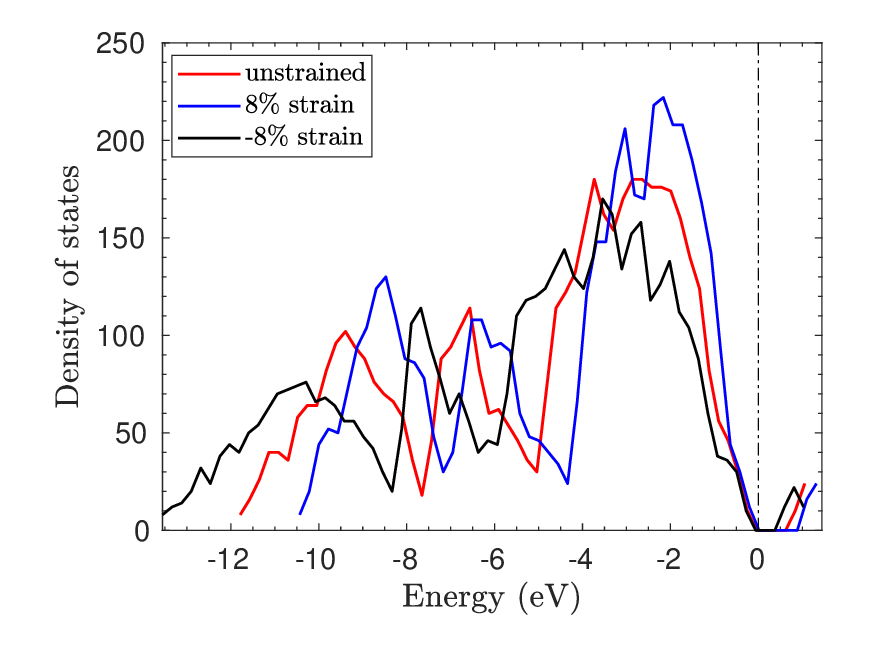}\label{Fig:DOSf}}
{\caption{Density of states of pristine, n-type doped with P and p-type doped with B silicon quantum dots of smallest size (Si$_{5}$H$_{12}$) in (a), (c) and (e), and largest size (Si$_{525}$H$_{276}$) (b), (d) and (f) is shown. The Fermi energy is shifted to zero and is shown as a black dash-dotted line.}\label{Fig:S:DOS}}
\end{figure}
Figure \ref{Fig:S:DOS} shows the density of states (DOS) for doped and strained Si$_{29}$H$_{36}$ and Si$_{525}$H$_{275}$ quantum dots. We consider doping at the center with P for n-type and B for p-type as representative examples. As evident from these plots, strain and doping significantly affect the density of states in both cases.  We see that the peaks of the DOS decrease with compressive strains and increase with tensile strains. The position of the peaks also changes with strain, where the energy level corresponding to the peaks is shifted to the right i.e. increases with tensile strain, and moved to the left i.e. decreases with compressive strain. 

\begin{figure}[H]\centering
\subfloat[pristine, -8\% strain]{\includegraphics[keepaspectratio=true,width=0.35\textwidth]{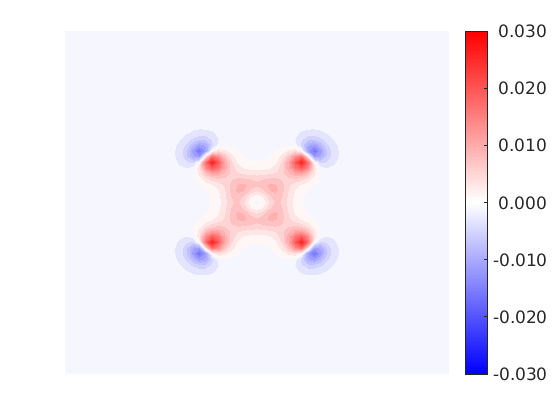}\label{Fig:Den:Da}}
\subfloat[pristine, -8\% strain]{\includegraphics[keepaspectratio=true,width=0.35\textwidth]{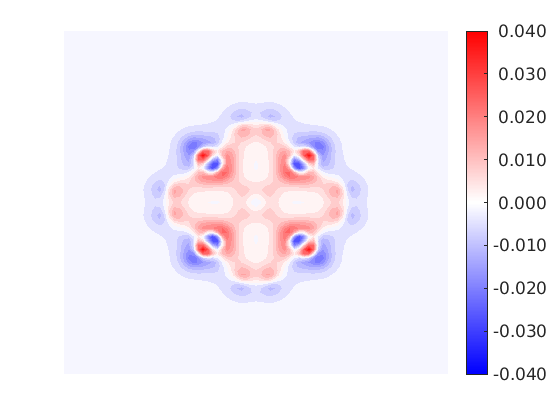}\label{Fig:Den:Db}}
\subfloat[pristine, -8\% strain]{\includegraphics[keepaspectratio=true,width=0.35\textwidth]{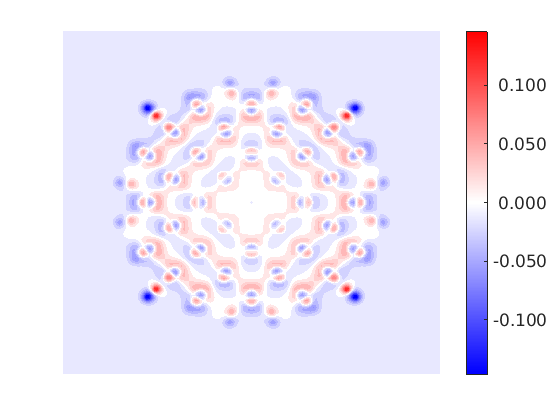}\label{Fig:Den:Dc}} \\
\subfloat[pristine, 8\% strain]{\includegraphics[keepaspectratio=true,width=0.35\textwidth]{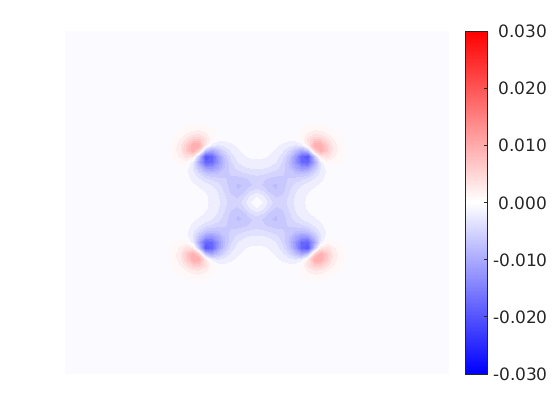}\label{Fig:Den:Dd}}
\subfloat[pristine, 8\% strain]{\includegraphics[keepaspectratio=true,width=0.35\textwidth]{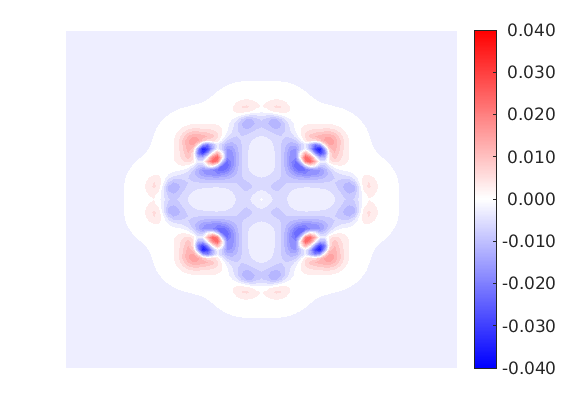}\label{Fig:Den:De}}
\subfloat[pristine, 8\% strain]{\includegraphics[keepaspectratio=true,width=0.35\textwidth]{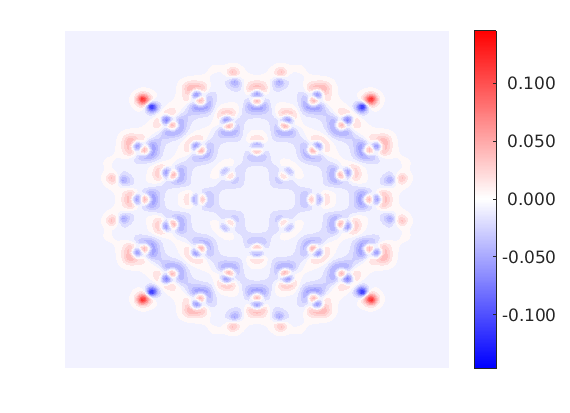}\label{Fig:Den:Df}}
{\caption{Electron density difference for 8 \% compression and 8 \% tension cases for Si$_{5}$H$_{12}$, Si$_{29}$H$_{36}$ and Si$_{525}$H$_{276}$. (a)-(c) shows compressive strains and (d) - (f) shows tensile strains.}\label{Fig:S:DEN}}
\end{figure}

Finally{,} we make some remarks on the electron density difference $\Delta \rho$. The electron density difference is defined as $\Delta \rho = \rho' - \rho_0$, where $\rho'$ is the electron density of the strained and/or doped quantum dot and $\rho_0$ {is} the electron density of the pristine unstrained quantum dot. By this definition, a positive electron density difference implies that the value of the electron density in the strained/doped configuration is greater than the value of the electron density in the pristine quantum dot. Figure \ref{Fig:S:DEN} shows the {effect} of strain on the electron density difference for the smallest (Si$_{5}$H$_{12}$), intermediate (Si$_{29}$H$_{36}$) and largest (Si$_{525}$H$_{276}$) quantum dot. We see that compressive strains increase the electron density near the core of the quantum dot, while tensile strains deplete the electron density. The electron density near the edge of the quantum dot is depleted by tensile strains and increased by compressive strains.

Changes in electron density influence the charge distribution and electronic band structure, and can be correlated to the optical and electronic properties of the quantum dots. Higher electron density in specific regions leads to stronger absorption in specific wavelength ranges. In quantum dots, the photons are absorbed on the surface, and thus the changes in density near the surface strongly affect the optoelectronic properties. From Figure \ref{Fig:S:DEN}, we see that in large quantum dots, compressive strains deplete the electron density near the surface while tensile strain enhances it. Thus{,} large quantum dots under tension are expected to display enhanced optoelectronic properties. For small-sized dots (such as Si$_{5}$H$_{12}$), the number density of silicon atoms near the surface is more than the interior. Therefore, compressive strains enhance the electron density near the surface while tensile strains deplete the surface electron density. Overall, these observations corroborate the trends of the energy gap discussed earlier in this work.

\begin{figure}[H]\centering
\subfloat[P doped]{\includegraphics[keepaspectratio=true,width=0.25\textwidth]{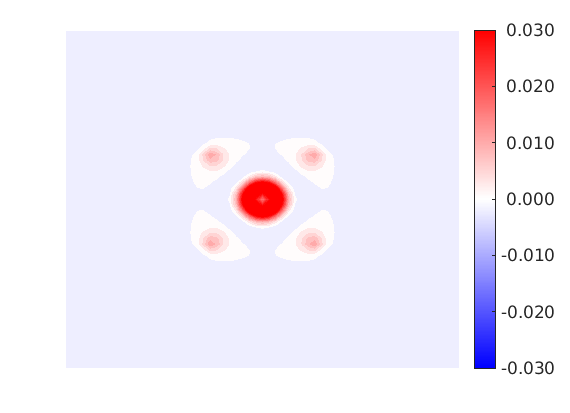}\label{Fig:Den:CE1}} 
\subfloat[B doped]{\includegraphics[keepaspectratio=true,width=0.25\textwidth]{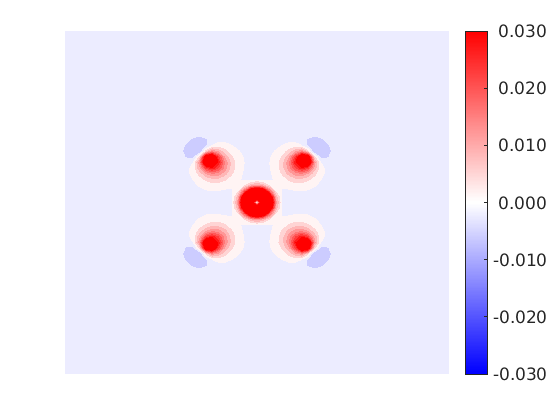}\label{Fig:Den:CE1}} 
\subfloat[As doped]{\includegraphics[keepaspectratio=true,width=0.25\textwidth]{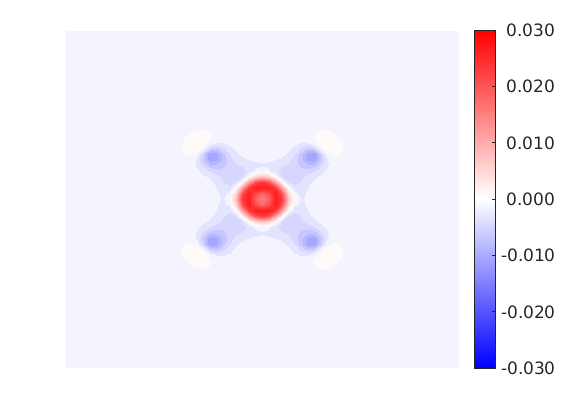}\label{Fig:Den:CE1}} 
\subfloat[Al doped]{\includegraphics[keepaspectratio=true,width=0.25\textwidth]{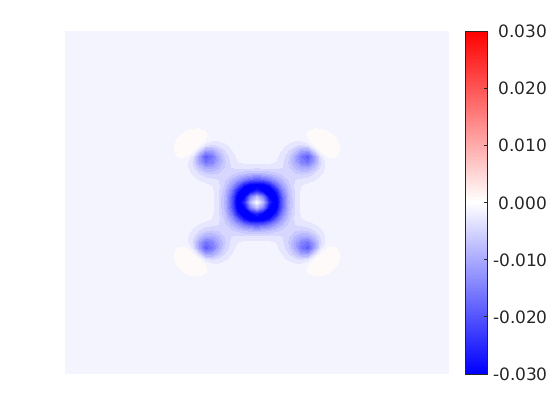}\label{Fig:Den:CE1}} \\
\subfloat[P doped]{\includegraphics[keepaspectratio=true,width=0.25\textwidth]{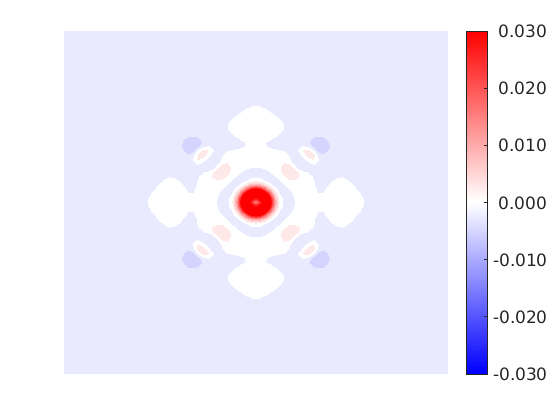}\label{Fig:Den:CE1}} 
\subfloat[B doped]{\includegraphics[keepaspectratio=true,width=0.25\textwidth]{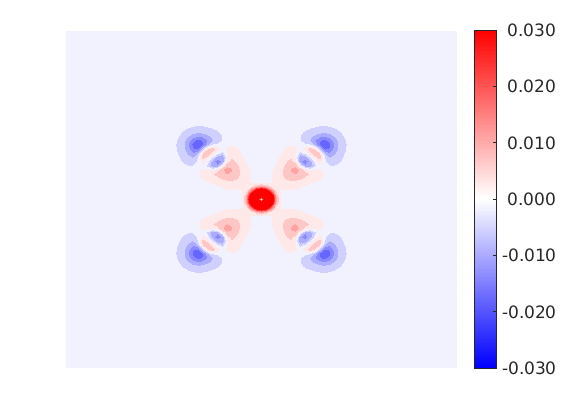}\label{Fig:Den:CE1}} 
\subfloat[As doped]{\includegraphics[keepaspectratio=true,width=0.25\textwidth]{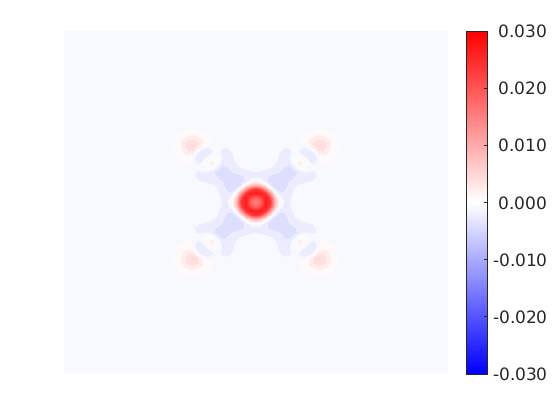}\label{Fig:Den:CE1}} 
\subfloat[Al doped]{\includegraphics[keepaspectratio=true,width=0.25\textwidth]{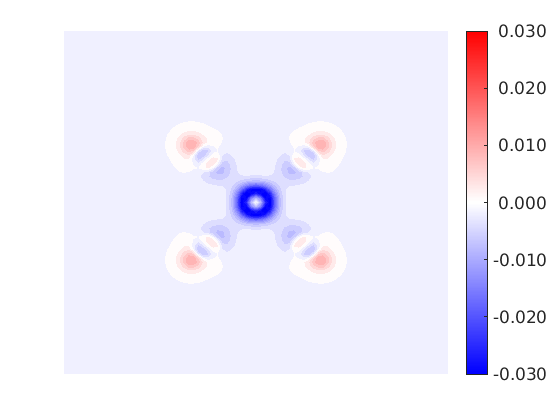}\label{Fig:Den:CE1}} \\
\subfloat[P doped]{\includegraphics[keepaspectratio=true,width=0.25\textwidth]{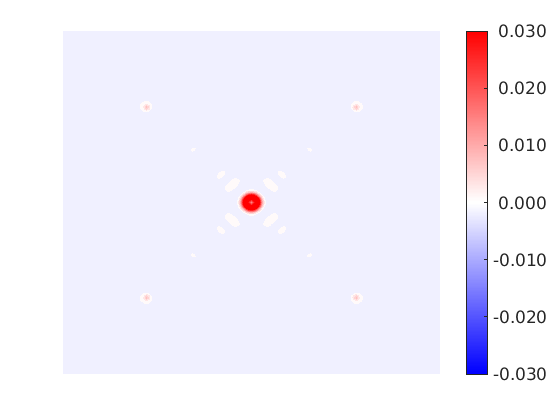}\label{Fig:Den:CE1}} 
\subfloat[B doped]{\includegraphics[keepaspectratio=true,width=0.25\textwidth]{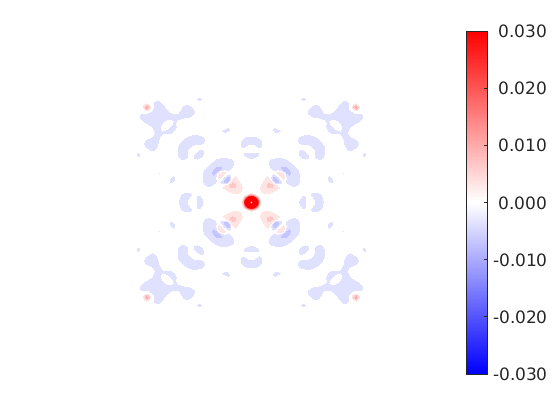}\label{Fig:Den:CE1}} 
\subfloat[As doped]{\includegraphics[keepaspectratio=true,width=0.25\textwidth]{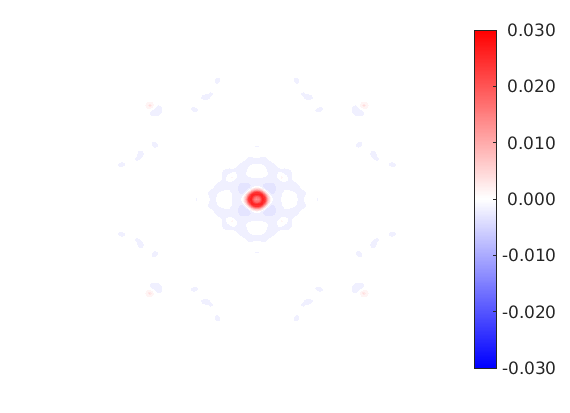}\label{Fig:Den:CE1}} 
\subfloat[Al doped]{\includegraphics[keepaspectratio=true,width=0.25\textwidth]{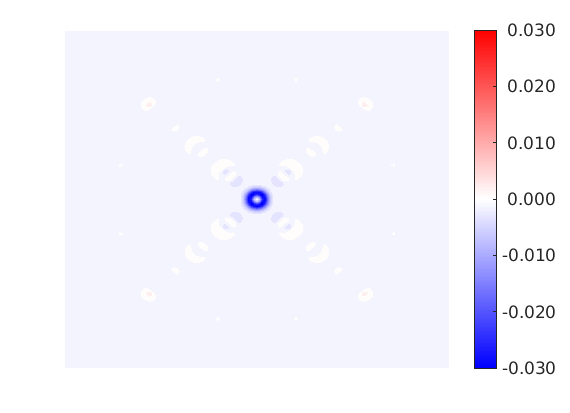}\label{Fig:Den:CE1}} 
{\caption{Electron density difference for doped quantum dots without strain. In (a)-(d) doped Si$_5$H$_{12}$, in (e)-(h) doped Si$_{29}$H$_{36}$ and in (i)-(l) doped Si$_{525}$H$_{276}$ is shown.}\label{Fig:D:DEN}}
\end{figure}

Figure \ref{Fig:D:DEN} shows the electron density difference due to dopant atoms in the absence of strain. Again we consider the smallest (Si$_{5}$H$_{12}$), an intermediate (Si$_{29}$H$_{36}$) and largest (Si$_{525}$H$_{276}$) quantum dots with dopants placed at the center are representative examples. We see that the perturbations due to B and Al atoms are long-ranged as seen in Figure \ref{Fig:D:DEN}. These oscillations in the electron density difference can be understood in terms of the misfit volumes ($\Delta V = V_{dopant}-V_{Si}$). The difference in the atomic volumes are $-1.39$, $-3.00$, $2.61$ and $0.80$ Angstrom$^{3}$ for P, B, As and Al dopants. The magnitude of misfit volume for B and Al atoms {is} largest{,} giving rise to the long-range oscillations in the electron density difference. 


\begin{figure}[H]\centering
\includegraphics[keepaspectratio=true,width=0.9\textwidth]{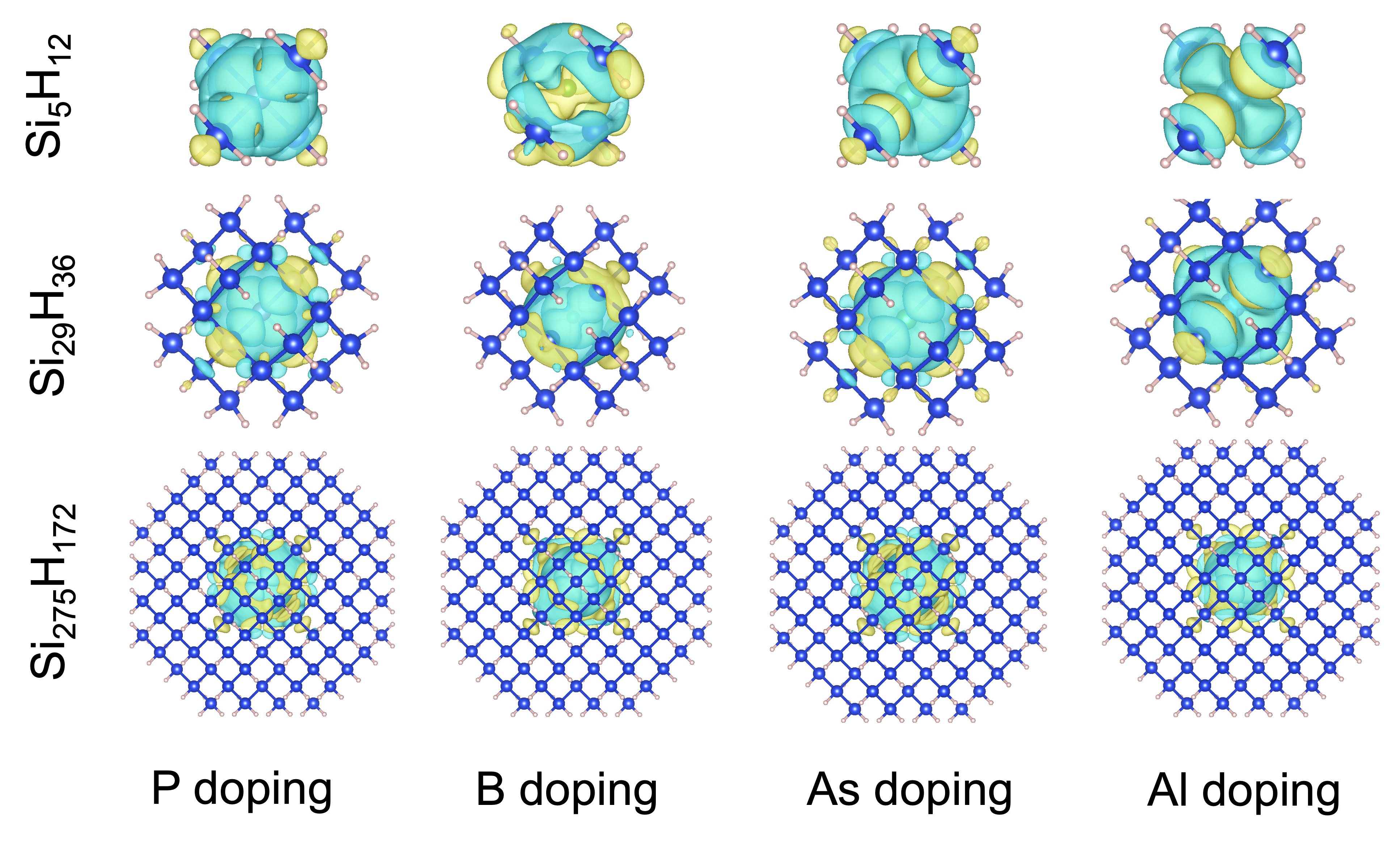}
\caption{{Charge density difference of small, intermediate, and large-sized quantum dots with P, B, As, and Al doping. Yellow and green colors denote positive and negative values of the charge density difference, respectively.}}\label{Fig:ChargeDifference}
\end{figure}

In Figure \ref{Fig:ChargeDifference} we plot the charge density difference of doped small, intermediate, and large quantum dots using the VESTA \cite{VESTA} software package. The charge density difference is defined as $\rho_{Si_{n-1}H_{m}X} - \rho_{Si_{n-1}H_{m}} - \rho_{X}$, where $\rho_{Si_{n-1}H_{m}X}$ is the charge density of Si$_{n-1}$H$_{m}$X in fully relaxed configuration, $\rho_{Si_{n-1}H_{m}}$ and $\rho_{X}$ are the charge density of Si$_{n-1}$H$_{m}$ quantum dot and the dopant $X$ with the same atomic coordinates as the relaxed configuration of Si$_{n-1}$H$_{m}$X quantum dot. {From Figure \ref{Fig:ChargeDifference}, we see that in small quantum dots, the charge density difference is delocalized in the entire volume of the dot. When the size of the quantum dot is increased from small to intermediate and then to large, the charge density difference is localized near the center, and its effect on the edge diminishes.} Dopants increase the concentration of charge carriers and thereby screen Coulomb interactions between electrons and holes, leading to changes in the energy gap. We also see that the ratio of the charge density difference to the quantum dot volume arising from a single dopant decreases with the increase in quantum dot size. This effectively leads to a reduction in Coulomb screening, and therefore its effect on {the} energy gap diminishes as observed in Figure \ref{Fig:BG}.

\section{Concluding remarks}
In this work, we have studied the effect of doping and strain on the stability, energy gap, Fermi level, electronic density, and density of states of silicon quantum dots. We have chosen quantum dots of sizes ranging from $\sim$ 0.6 nanometers to $\sim$ 3 nanometers. We have considered both n- and p-type doping of silicon quantum dots, where P and As atoms are used for n-type doping and B and Al atoms are used for p-type doping. Our work shows that increasing quantum dot size decreases cohesive energy and the energy gap. The cohesive energy and energy gap are strongly {affected} by strain. Furthermore, the response to strain also depends on the size of the quantum dot and {the} dopant type. To capture the dependence of quantum dot size and strain, we have presented expressions of cohesive energy and energy gap as power-law of size and {a} polynomial dependence on strain. We have also shown that the Fermi energy increases with size for pristine and p-type doping but decreases with size for n-type doping. Finally, our calculations have also captured the effect of strain and dopant type on the density of states and electron density of the quantum dots. 

Overall, the computational study presented in this work can be used to guide experiments and device optimizations. Specifically, while doped quantum dots are widely used in photovoltaic applications, our results show that additional strain can be used to tune the energy gap. In practice, strain on the nanocluster can be imparted in several ways{,} such as nanoclusters deposited on a strained substrate or encased in a strained matrix.

This study opens the scope for several possible future investigations on silicon quantum dots. Here we mention some of these. In this work{,} we have considered doping by a single dopant at a site in the nanocluster. An interesting avenue of investigation is studying the influence of doping by more than one type of dopants at multiple sites and studying the effects of distribution of dopants, dopant interaction \cite{Ghosh:SQ}, short range ordering \cite{Ghosh:SRO}, strain and temperature \cite{Ghosh:MgAl} on the energetics and electronic properties of nanoclusters. In this work, we have considered silicon quantum dots without defects. It is therefore natural to raise the question of the effect of defects on the electronic properties of these quantum dots. This is important when considering large tensile strains{,} which can nucleate defects \cite{ghosh:2019}. Specifically, questions regarding nucleating defects interacting with dopants and their effects on the electronic properties arise. 
\section*{Data availability}
The datasets used is available from the corresponding author on reasonable request.

\bibliographystyle{unsrt}
\bibliography{Manuscript}

\begin{thebibliography}{10}

\bibitem{soga:2006}
Tetsuo Soga.
\newblock {\em Nanostructured materials for solar energy conversion}.
\newblock Elsevier, 2006.

\bibitem{wheeler:2013}
Lance~M Wheeler, Nathan~R Neale, Ting Chen, and Uwe~R Kortshagen.
\newblock Hypervalent surface interactions for colloidal stability and doping
  of silicon nanocrystals.
\newblock {\em Nature communications}, 4(1):2197, 2013.

\bibitem{warner:2005}
Jamie~H Warner, Akiyoshi Hoshino, Kenji Yamamoto, and Richard~D Tilley.
\newblock Water-soluble photoluminescent silicon quantum dots.
\newblock {\em Angewandte Chemie International Edition}, 44(29):4550--4554,
  2005.

\bibitem{cho:2007}
Eun-Chel Cho, Martin~A Green, Gavin Conibeer, Dengyuan Song, Young-Hyun Cho,
  Giuseppe Scardera, Shujuan Huang, Sangwook Park, XJ~Hao, Yidan Huang, et~al.
\newblock Silicon quantum dots in a dielectric matrix for all-silicon tandem
  solar cells.
\newblock {\em Advances in OptoElectronics}, 2007(1):069578, 2007.

\bibitem{peng:2006}
X-H Peng, S~Ganti, A~Alizadeh, P~Sharma, SK~Kumar, and SK~Nayak.
\newblock Strain-engineered photoluminescence of silicon nanoclusters.
\newblock {\em Physical Review B}, 74(3):035339, 2006.

\bibitem{canham:1990}
Leigh~T Canham.
\newblock Silicon quantum wire array fabrication by electrochemical and
  chemical dissolution of wafers.
\newblock {\em Applied physics letters}, 57(10):1046--1048, 1990.

\bibitem{delley:1993}
B~Delley and EF~Steigmeier.
\newblock Quantum confinement in si nanocrystals.
\newblock {\em Physical Review B}, 47(3):1397, 1993.

\bibitem{wang:1994}
Lin~Wang Wang and Alex Zunger.
\newblock Electronic structure pseudopotential calculations of large (. apprx.
  1000 atoms) si quantum dots.
\newblock {\em The Journal of Physical Chemistry}, 98(8):2158--2165, 1994.

\bibitem{Proot:1992}
JP~Proot, C~Delerue, and G~Allan.
\newblock Electronic structure and optical properties of silicon crystallites:
  Application to porous silicon.
\newblock {\em Applied Physics Letters}, 61(16):1948--1950, 1992.

\bibitem{delerue:2000}
Christophe Delerue, Michel Lannoo, and Guy Allan.
\newblock Excitonic and quasiparticle gaps in si nanocrystals.
\newblock {\em Physical review letters}, 84(11):2457, 2000.

\bibitem{ougut:1997}
Serdar {\"O}{\u{g}}{\"u}t, James~R Chelikowsky, and Steven~G Louie.
\newblock Quantum confinement and optical gaps in si nanocrystals.
\newblock {\em Physical Review Letters}, 79(9):1770, 1997.

\bibitem{duan:2015}
Jialong Duan, Huihui Zhang, Qunwei Tang, Benlin He, and Liangmin Yu.
\newblock Recent advances in critical materials for quantum dot-sensitized
  solar cells: a review.
\newblock {\em Journal of Materials Chemistry A}, 3(34):17497--17510, 2015.

\bibitem{oliva:2016}
Brittany~L Oliva-Chatelain, Thomas~M Ticich, and Andrew~R Barron.
\newblock Doping silicon nanocrystals and quantum dots.
\newblock {\em Nanoscale}, 8(4):1733--1745, 2016.

\bibitem{pi:2011}
Xiaodong Pi, Qing Li, Dongsheng Li, and Deren Yang.
\newblock Spin-coating silicon-quantum-dot ink to improve solar cell
  efficiency.
\newblock {\em Solar Energy Materials and Solar Cells}, 95(10):2941--2945,
  2011.

\bibitem{pi:2012}
Xiaodong Pi.
\newblock Doping silicon nanocrystals with boron and phosphorus.
\newblock {\em Journal of Nanomaterials}, 2012:3--3, 2012.

\bibitem{huang:2012}
Shujuan Huang, Yong~Heng So, Gavin Conibeer, and Martin Green.
\newblock Doping of silicon quantum dots embedded in nitride matrix for
  all-silicon tandem cells.
\newblock {\em Japanese Journal of Applied Physics}, 51(10S):10NE10, 2012.

\bibitem{hao:2009}
XJ~Hao, E-C Cho, Giuseppe Scardera, YS~Shen, Edith Bellet-Amalric, D~Bellet,
  Gavin Conibeer, and Martin~Andrew Green.
\newblock Phosphorus-doped silicon quantum dots for all-silicon quantum dot
  tandem solar cells.
\newblock {\em Solar Energy Materials and Solar Cells}, 93(9):1524--1530, 2009.

\bibitem{hao:2009b}
XJ~Hao, Eun-Chel Cho, Christopher Flynn, YS~Shen, Sung-Chan Park, Gavin
  Conibeer, and Martin~Andrew Green.
\newblock Synthesis and characterization of boron-doped si quantum dots for
  all-si quantum dot tandem solar cells.
\newblock {\em Solar Energy Materials and Solar Cells}, 93(2):273--279, 2009.

\bibitem{hao:2008}
XJ~Hao, EC~Cho, C~Flynn, YS~Shen, G~Conibeer, and MA~Green.
\newblock Effects of boron doping on the structural and optical properties of
  silicon nanocrystals in a silicon dioxide matrix.
\newblock {\em Nanotechnology}, 19(42):424019, 2008.

\bibitem{perez:2009}
Ivan Perez-Wurfl, Xiaojing Hao, Angus Gentle, Dong-Ho Kim, Gavin Conibeer,
  Martin Green, et~al.
\newblock Si nanocrystal pin diodes fabricated on quartz substrates for third
  generation solar cell applications.
\newblock {\em Applied Physics Letters}, 95(15), 2009.

\bibitem{pagliaro:2008}
Mario Pagliaro, Rosaria Ciriminna, and Giovanni Palmisano.
\newblock Flexible solar cells.
\newblock {\em ChemSusChem: Chemistry \& Sustainability Energy \& Materials},
  1(11):880--891, 2008.

\bibitem{sze:2008}
Simon~Min Sze.
\newblock {\em Semiconductor devices: physics and technology}.
\newblock John wiley \& sons, 2008.

\bibitem{ossicini:2005}
Stefano Ossicini, Elena Degoli, F~Iori, E~Luppi, Rita Magri, G~Cantele,
  F~Trani, and Domenico Ninno.
\newblock Simultaneously b-and p-doped silicon nanoclusters: Formation energies
  and electronic properties.
\newblock {\em Applied Physics Letters}, 87(17), 2005.

\bibitem{zhang:2017}
Hui Zhang, Runmin Zhang, Katelyn~S Schramke, Nicholas~M Bedford, Katharine
  Hunter, Uwe~R Kortshagen, and Peter Nordlander.
\newblock Doped silicon nanocrystal plasmonics.
\newblock {\em Acs Photonics}, 4(4):963--970, 2017.

\bibitem{delley:1995}
B~Delley and EF~Steigmeier.
\newblock Size dependence of band gaps in silicon nanostructures.
\newblock {\em Applied physics letters}, 67(16):2370--2372, 1995.

\bibitem{buda:1992}
F~Buda, J~Kohanoff, and M~Parrinello.
\newblock Optical properties of porous silicon: A first-principles study.
\newblock {\em Physical review letters}, 69(8):1272, 1992.

\bibitem{melnikov:2003}
Dmitriy~V Melnikov and James~R Chelikowsky.
\newblock Absorption spectra of germanium nanocrystals.
\newblock {\em Solid state communications}, 127(5):361--365, 2003.

\bibitem{hohenberg:1964}
Pierre Hohenberg and Walter Kohn.
\newblock Inhomogeneous electron gas.
\newblock {\em Physical review}, 136(3B):B864, 1964.

\bibitem{kohn:1965}
Walter Kohn and Lu~Jeu Sham.
\newblock Self-consistent equations including exchange and correlation effects.
\newblock {\em Physical review}, 140(4A):A1133, 1965.

\bibitem{ghosh:2017a}
Swarnava Ghosh and Phanish Suryanarayana.
\newblock Sparc: Accurate and efficient finite-difference formulation and
  parallel implementation of density functional theory: Isolated clusters.
\newblock {\em Computer Physics Communications}, 212:189--204, 2017.

\bibitem{ghosh:2017b}
Swarnava Ghosh and Phanish Suryanarayana.
\newblock Sparc: Accurate and efficient finite-difference formulation and
  parallel implementation of density functional theory: Extended systems.
\newblock {\em Computer Physics Communications}, 216:109--125, 2017.

\bibitem{troullier:1991}
Norman Troullier and Jos{\'e}~Lu{\'\i}s Martins.
\newblock Efficient pseudopotentials for plane-wave calculations.
\newblock {\em Physical review B}, 43(3):1993, 1991.

\bibitem{perdew:1992}
John~P Perdew and Yue Wang.
\newblock Accurate and simple analytic representation of the electron-gas
  correlation energy.
\newblock {\em Physical review B}, 45(23):13244, 1992.

\bibitem{ceperley:1980}
David~M Ceperley and Berni~J Alder.
\newblock Ground state of the electron gas by a stochastic method.
\newblock {\em Physical review letters}, 45(7):566, 1980.

\bibitem{baerends:2013}
EJ~Baerends, OV~Gritsenko, and R~Van~Meer.
\newblock The kohn--sham gap, the fundamental gap and the optical gap: the
  physical meaning of occupied and virtual kohn--sham orbital energies.
\newblock {\em Physical Chemistry Chemical Physics}, 15(39):16408--16425, 2013.

\bibitem{VESTA}
Koichi Momma and Fujio Izumi.
\newblock Vesta: a three-dimensional visualization system for electronic and
  structural analysis.
\newblock {\em Applied Crystallography}, 41(3):653--658, 2008.

\bibitem{nogueira2003tutorial}
Fernando Nogueira, Alberto Castro, and Miguel~AL Marques.
\newblock A tutorial on density functional theory.
\newblock {\em A primer in density functional theory}, pages 218--256, 2003.

\bibitem{smith:2005}
A~Smith, ZH~Yamani, N~Roberts, J~Turner, SR~Habbal, S~Granick, and MH~Nayfeh.
\newblock Observation of strong direct-like oscillator strength in the
  photoluminescence of si nanoparticles.
\newblock {\em Physical Review B—Condensed Matter and Materials Physics},
  72(20):205307, 2005.

\bibitem{wilcoxon:1999}
JP~Wilcoxon, GA~Samara, and PN~Provencio.
\newblock Optical and electronic properties of si nanoclusters synthesized in
  inverse micelles.
\newblock {\em Physical Review B}, 60(4):2704, 1999.

\bibitem{Ghosh:SQ}
Swarnava Ghosh and Kaushik Bhattacharya.
\newblock Spectral quadrature for the first principles study of crystal
  defects: Application to magnesium.
\newblock {\em Journal of Computational Physics}, 456:111035, 2022.

\bibitem{Ghosh:SRO}
Gautam Anand, Swarnava Ghosh, and Markus Eisenbach.
\newblock Order parameter engineering for random systems.
\newblock {\em High Entropy Alloys \& Materials}, 1(2):271--284, 2023.

\bibitem{Ghosh:MgAl}
Swarnava Ghosh and Kaushik Bhattacharya.
\newblock Influence of thermomechanical loads on the energetics of
  precipitation in magnesium aluminum alloys.
\newblock {\em Acta Materialia}, 193:28--39, 2020.

\bibitem{ghosh:2019}
Swarnava Ghosh and Phanish Suryanarayana.
\newblock Electronic structure study regarding the influence of macroscopic
  deformations on the vacancy formation energy in aluminum.
\newblock {\em Mechanics Research Communications}, 99:58--63, 2019.

\end{thebibliography}
\section*{Acknowledgements}
We gratefully acknowledge the valuable suggestions of the anonymous reviewers. This research used resources of the Oak Ridge Leadership Computing Facility, a DOE Office of Science User Facility operated by the Oak Ridge National Laboratory under contract DE-AC05-00OR22725. This research also used resources of the Compute and Data Environment for Science (CADES) at the Oak Ridge National Laboratory, which is supported by
the Office of Science of the U.S. Department of Energy under Contract No. DE-AC05-00OR22725.
This manuscript has been authored in part by UT-Battelle, LLC, under contract DE-AC05- 00OR22725 with the US Department of Energy (DOE). The publisher acknowledges the US government license to provide public access under the DOE Public Access Plan (http://energy.gov/downloads/doe- public-access-plan).

\section*{Author contributions}
S.G. and M.E. conceived the project, S.G. performed simulations and analysis, S.G. and M.E. participated in discussions, S.G. wrote the draft, S.G. and M.E revised the draft.  
\section*{Competing interests}
The authors declare no competing interests.



\end{document}